\documentclass[a4paper,fleqn,usenatbib]{mnras}
\usepackage[]{natbib, graphicx, float, appendix, threeparttable}
\usepackage[FIGTOPCAP]{subfigure}
\usepackage[section]{placeins}
\usepackage{wrapfig}
\usepackage[fleqn]{amsmath}
\usepackage{xcolor}
\def\ga{{\rm\thinspace gauss}}

\def\Lsun{\hbox{$\rm\thinspace L_{\odot}$}}

\def\Msun{\hbox{$\rm\thinspace M_{\odot}$}}

\newcommand{\RNum}[1]{\uppercase\expandafter{\romannumeral #1\relax}}

\title[2MASX J00423991+3017515: An offset active galactic nucleus]{2MASX J00423991+3017515: An offset active galactic nucleus in an interacting system.}

\author[J. Drew Hogg et al.]{J.~Drew~Hogg$^{1,2}$\thanks{E-mail: drewhogg@astro.umd.edu}, 
Laura~Blecha$^{3}$, 
Christopher~S.~Reynolds$^{4}$,
Krista Lynne Smith$^{5}$,\newauthor
and Lisa M. Winter$^{6}$
\\
$^{1}$Department of Astronomy, University of Maryland, College Park, MD 20742, USA\\
$^{2}$Atos zData, Newark, DE 19711, USA\\
$^{3}$Physics Department, University of Florida, Gainesville, FL 32611, USA\\
$^{4}$Institute of Astronomy, University of Cambridge, Madingley Road, Cambridge. CB3 0HA, UK\\
$^{5}$Department of Physics, Southern Methodist University, Dallas, TX 75205, USA\\
$^{6}$Division of Atmospheric and Geospace Sciences, National Science Foundation, Alexandria, VA 22314, USA\\
}

\begin{document}
\label{firstpage}
\pagerange{\pageref{firstpage}--\pageref{lastpage}}
\maketitle

\begin{abstract}

We present a spectroscopic and imaging study of an abnormal active galactic nucleus (AGN), 2MASX J00423991+3017515.  This AGN is newly identified in the hard X-rays by the \emph{Swift} BAT All-Sky survey and found in an edge-on disk galaxy interacting with a nearby companion.  Here, we analyze the first optical spectra obtained for this system (taken in 2011 and 2016), high-resolution imaging taken with the \emph{Hubble Space Telescope} and \emph{Chandra} X-ray Observatory, and $1\arcsec$ imaging with the \emph{Very Large Array}.  Two unique properties are revealed: the peaks of the broad Balmer emission lines (associated with gas orbiting very near the supermassive black hole) are blue shifted from the corresponding narrow line emission and host galaxy absorption by $1540$ km s$^{-1}$, and the AGN is spatially displaced from the apparent center of its host galaxy by $3.8$ kpc.  We explore several scenarios to explain these features, along with other anomalies, and propose that 2MASX J00423991+3017515 may be an AGN with an unusually strong wind residing in a uniquely configured major merger, or that it is an AGN recoiling from either a gravitational ``slingshot" in a three-body interaction or from a kick due to the asymmetric emission of gravitational waves following the coalescence of two progenitor supermassive black holes.
 
\end{abstract}


\section{Introduction}
\label{sec-intro}

The hierarchical assembly of massive galaxies through the successive merging of smaller galaxies is a core tenet of our galaxy formation paradigm \citep{1978MNRAS.183..341W}.  These mergers are dynamic events that can grow bulges, alter galaxy morphologies, and cause bursts of star formation.  Additionally, galaxies are known to host supermassive black holes (SMBHs), and mergers mark significant periods in their growth as well.  As pre-merger galaxies interact and merge, torques on their interstellar media can lead to accretion as material is funneled to the SMBHs \citep{1996ApJ...464..641M} which triggers AGN activity \citep[e.g.,][]{2012ApJ...758L..39T, 2014ApJ...789..112C, 2017ApJ...838..129B, 2018Natur.563..214K, 2018MNRAS.476.2308W}.  Galaxy mergers should also lead to the formation of SMBH binaries, which produce powerful gravitational waves \citep[GWs,][]{Einstein:1916cc, 1918SPAW.......154E} if they inspiral and eventually merge.

SMBH binary coalescence depends on the interplay of several physical mechanisms and the timescales are not well constrained.  As the postmerger remnant galaxy relaxes, the SMBHs from the pre-merger hosts will inevitably settle to the bottom of the gravitational potential due to dynamical friction and form a bound binary \citep{1980Natur.287..307B}.  Further binary inspiral is driven by interactions with nearby stars and gas that may either shrink the binary orbit efficiently or cause the inspiral to ``stall" for many Gyr \cite[e.g.,][]{2003AIPC..686..201M, 2006ApJ...642L..21B, 2013ApJ...773..100K, 2017MNRAS.471.4508K}. Eventually, at $\sim$mpc scales, the emission of GWs will dominate the inspiral of the SMBH binary and efficiently drive the SMBHs to coalescence.

Directly measuring the GWs mediating an inspiral is the clearest way to detect coalescing SMBHs, but to date, they have not been detected.  In the lower mass regime, the detection of GWs emitted during the coalescence of a $35\:\Msun$ black hole with a $30\:\Msun$ black hole in the GW150914 event \citep{2016PhRvL.116f1102A} provided the first direct evidence that black holes can indeed grow through this pathway.  Since then, numerous other GW events have been recorded from the coalescence of different species of compact objects, including the GW190521 event which produced an intermediate mass black hole of mass $\sim$150 $\Msun$ \citep{PhysRevLett.125.101102}.  However, GWs from SMBH binary systems are produced at lower frequencies (nHz-mHz). SMBH mergers could be directly detected with a space-based GW detector such as LISA \citep{2017arXiv170200786A}, and mpc-scale SMBH binaries should be the main contributor to the GW background detectable with pulsar timing arrays (PTAs). PTAs monitor millisecond pulsars to search for coordinated shifts in pulse arrival times caused by nHz GWs \citep{1990ApJ...361..300F, 2008MNRAS.390..192S}. While no detection has yet been made, current upper limits and theoretical predictions suggest that the stochastic GW background should be detected in the very near future \citep{2015Sci...349.1522S, 2016ApJ...821...13A, 2016MNRAS.458.1267V, 2016ApJ...819L...6T, 2017MNRAS.471.4508K, 2020AAS...23611107S}.

A complementary means of probing SMBH binary inspiral and merger is to search for recoiling AGNs.  When the two progenitor, inspiraling SMBHs have unequal masses or they are spinning, the GW emission they produce will be asymmetric.  At coalescence, this will impart linear momentum to the newly formed SMBH and cause it to recoil from the ``kick" \citep{1962PhRv..128.2471P, 1973ApJ...183..657B}.  If the binary is actively accreting at the time of coalescence, it should continue to do so after.  The resulting recoiling AGN should retain bound gas that has an orbital velocity greater than the kick velocity; a condition that is satisfied within a volume whose radius encompasses the accretion disk and broad line region (BLR) for a reasonable range of recoil kicks.  With this material, a recoiling AGN can continue to accrete for tens of Myr, but its trajectory depends on many factors such as gas drag and even the timing of the SMBH merger relative to the rapidly-changing depth of the host gravitational potential well \citep{2011MNRAS.412.2154B}.  Nonetheless, theoretical models predict that a population of offset, recoiling AGN should be detectable \citep{2008ApJ...687L..57V, 2016MNRAS.456..961B}.  Recoiling AGNs could be identifiable by two main distinguishing properties: a velocity offset in the peaks of the broad Balmer line emission, and a spatial displacement of the AGN from the dynamical center, and presumably the visual center, of the host galaxy \citep[e.g.,][]{2004ApJ...606L..17M, 2007PhRvL..99d1103L}.  Additionally, other telling signatures may be present including tidal tails or other morphological disturbances indicative of a recent major merger.

However, since the timescales associated with the process are not well known, a direct path to inspiral of a SMBH binary is not a foregone conclusion. For example, the initial shrinking of a SMBH binary occurs when stars in the so-called ``loss-cone" of low-angular-momentum stellar orbits are scattered.  Eventually, the loss cone can be depleted and if it is only replenished via two-body relaxation, which is a slow diffusive process, the binary shrinkage may stall and significantly increase the inspiral timescale \citep[the ``final parsec problem,"][]{2001ApJ...563...34M}.  When the inspiral timescale is longer than the time between galaxy mergers, a third SMBH can be added to the galaxy in a subsequent galaxy merger.  This third SMBH will sink to the bottom of the gravitational potential and eventually encounter the stalled binary. Three-body interactions will typically eject the least massive SMBH, giving it a ``slingshot" kick, and the perturbation will greatly reduce the merger timescale for the two remaining SMBHs \citep{2007MNRAS.377..957H, 2016MNRAS.461.4419B, 2018MNRAS.477.3910B}.  The kicked SMBH would appear as a recoiling AGN if it is actively accreting, and could be difficult to distinguish from a GW recoiling AGN.  It is worth noting that the galaxy nucleus will be left with two SMBHs, and potentially two AGNs.

Positively and unambiguously identifying a genuine recoiling AGN is difficult because not only are these rare events, but there are several sources of confusion that make it hard to discern them from a sea of interlopers.  First, recoiling AGNs must be filtered from the standard AGN population.  As detailed multi-wavelength and time domain studies of individual AGN emerge, we are learning that ``typical," i.e., non-recoiling, AGN behavior is diverse with each object having its own peculiarities and idiosyncrasies.  For example, interpreting AGN broad line emission is notoriously tricky as it regularly changes shape and has been observed to even have multiple peaks, e.g., \citet{2007ApJS..169..167G} and \citet{2010ApJS..187..416L}.  Winds, and outflows in general, are also common from AGNs and can leave a variety of spectral imprints like shifted peaks in the broad line emission due to the bulk velocity of the outflowing gas.  Some AGNs have even changed spectral type \citep{2003MNRAS.342..422M, 2015ApJ...800..144L} or suddenly turned on \citep{2017ApJ...835..144G}.  Separating non-recoiling AGNs from recoiling AGNs can be further complicated if the AGN host galaxy has any structural artifacts from a recent major merger, as a ``typical" AGN could appear spatially displaced from such an event, rather than by its own motion.

\begin{table*}
	\centering
	\caption{Summary of Optical Spectroscopic Observation}
    \label{table-optical_spec}
	\begin{tabular}{ccccccccc} 
		\hline
        Date & Observatory &  $n_{exp}$ & Exp. Time & Grating(s) & Dispersion(s) & Slit Width  & Seeing & Airmass \\
		\hline
    UT 2011-08-07 & APO & $2$ & $1440$ & $400$, $300$ g/mm & $\Delta \lambda_{b}, \Delta \lambda_{r} = 1.83, 2.31\:\AA$/pix & $1.5\arcsec$ & $1.0\arcsec$ & 1.02\\
    UT 2016-12-05 & LDT & $3$ & $5400$ & $500$ g/mm & $\Delta \lambda = 1.3\:\AA$/pix & $1.5\arcsec$ & $1.5\arcsec$ & 1.04\\

		\hline
	\end{tabular}
\end{table*}

Second, AGNs in different stages of a galaxy merger can masquerade as a recoiling AGN and mimic signatures we would expect from these sources.  In particular, this means distinguishing recoiling AGNs from other unusual activities that might occur like galaxies that contain multiple SMBHs where only one is accreting. In rare cases, the recoil scenario may also be difficult to distinguish from a highly unusual supernova \citep{2014MNRAS.445..515K}.

In this paper, we investigate an AGN, 2MASX J00423991+3017515, whose broad Balmer emission lines display a large kinematic offset ($1540$ km s$^{-1}$) that is stable over five and a half years.  Imaging shows that the AGN host galaxy is an edge-on disk that is currently interacting with a nearby companion in the early stages of a merger.  From modeling the 2D light distribution of the system, it appears the AGN itself is displaced from the center of the host galaxy by several kiloparsecs.  The acquisition, reduction, and processing of the multi-wavelength spectroscopic and imaging data we collected are presented in Section \ref{sec-disc_obs}, the data analysis is described in Section \ref{sec-results}, a discussion of our results is given in Section \ref{sec-discussion}, and our conclusions are summarized in Section \ref{sec-conc}.

\section{Observations and Data Analysis}
\label{sec-disc_obs}

\subsection{Initial Discovery, Parent Sample, \& Overview of Observations}
\label{sec-disc_samp_goals}

2MASX J00423991+3017515 was serendipitously discovered while conducting the first optical spectroscopic follow-ups of newly detected AGNs in the Burst Alert Telescope (BAT) 70- and 105-month all-sky monitoring surveys \citep{2013ApJS..207...19B, 2018ApJS..235....4O}.  The BAT is one of three telescopes on the Neil Gehrels \emph{Swift} Observatory alongside the X-Ray Telescope (XRT) and the UV Optical Telescope (UVOT).  It is sensitive to hard X-ray emission, 14-195 keV, where radiation is less affected by the presence of intervening material, so the AGN population detected in the survey is nearly unbiased to the presence of surrounding dust and gas, at least up to Compton thicknesses (10$^{24}$ cm$^{-2}$).  Significant efforts have been dedicated to studying this sample of AGNs which has enriched our understanding of AGN demographics \citep{2009ApJ...690.1322W, 2010ApJ...710..503W, 2017ApJ...850...74K, 2017MNRAS.470..800T, 2017ApJS..233...17R}, the properties of AGN hosting galaxies \citep{2011ApJ...739...57K}, and AGN variability \citep{2009ApJ...701.1644W, 2013ApJ...770...60S}.  The sample has also uncovered numerous odd AGNs including X-ray bright, optically normal galaxies \citep{2014ApJ...794..112S}, AGNs that have a mismatch between their optical and X-ray classifications \citep{2012ApJ...752..153H}, and candidate recoiling AGNs \citep{2014MNRAS.445..515K}.  2MASX J00423991+3017515 adds to the collection of strange AGNs in the BAT all-sky survey.

\begin{table*}
	\centering
	\caption{Summary of HST Imaging}
	\label{table-hst_obs}
	\begin{tabular}{cccccc} 
		\hline
    Dataset ID & $n_{exp}$ & Total Exp. Time & Filter & Central Wavelength & Filter Width \\
		\hline
    IDA801010 & $2$ & $120$ & F814W & $8052$ \AA & $1536$ \AA\\
    IDA801020 & $2$ & $600$ & F814W & $8052$ \AA & $1536$ \AA\\
    IDA801030 & $2$ & $1440$ & F814W & $8052$ \AA & $1536$ \AA\\
    IDA801040 & $2$ & $60$ & F547M & $5447$ \AA & $650$ \AA\\
    IDA801050 & $2$ & $480$ & F547M & $5447$ \AA & $650$ \AA\\
    IDA801060 & $2$ & $960$ & F547M & $5447$ \AA & $650$ \AA\\
    IDA801070 & $3$ & $360$ & FQ750N & $7502$ \AA & $70$ \AA\\
    IDA801080 & $3$ & $1800$ & F225W & $2373$ \AA & $467$ \AA\\
		\hline
	\end{tabular}
\end{table*}

In this initial study of 2MASX J00423991+3017515, we obtained multi-wavelength spectroscopic and imaging data to delve into the peculiarities of the structure of the system and the nature of the kinematic offset in the broad Balmer lines.  The first optical spectrum was taken with the Dual Imaging Spectrograph (DIS) on the \emph{Apache Point Observatory} (APO) 3.5-meter telescope on UT 2011-08-07.  Additional spectroscopic observations with the DeVeny spectrograph on the 4.3-meter \emph{Lowell Discovery Telescope} (LDT) were attempted on UT 2016-10-08, UT 2016-12-05, UT 2017-10-13,  and UT 2017-12-08, but only data from UT 2016-12-05 was usable owing to adverse conditions on the other nights.  Note, though, that even on UT 2016-12-05, the conditions were non-photometric due to intermittent thin clouds and seeing that was comparable to the collimator slit width.  Nevertheless, this still provides a five-year baseline to monitor the broad line shapes and offsets.  Inspection of the system's Sloan Digital Sky Survey image showed what appeared to be a disturbed morphology.  This motivated \emph{Hubble Space Telescope} (HST) imaging with the goals of understanding the structure of the stellar component of the galaxy, locating the AGN or AGNs in system, understanding the distribution of narrow line emitting gas, and localizing the origin of the broadened H$\alpha$ emission.  X-ray observations with \emph{Chandra} were also taken to supplement the 0.2-10 keV spectra from XRT, and to aid in locating the primary AGN and searching for any additional AGN, if they might exist.  A short radio observation with the JVLA was obtained to help locate any heavily obscured AGNs.

\subsection{Optical Spectra}

Table \ref{table-optical_spec} provides the observational details of our optical spectroscopic data acquisition.  With both APO and LDT, two sets of spectra were taken to balance the need for broad wavelength coverage with the need to have good resolution.  One set had a grating tilt to capture the bluer end of the spectrum and the other had a grating tilt to capture the redder end.  All galaxy spectra were taken with the collimator slit oriented along the AGN hosting galaxy. The spectra taken on UT 2011-08-07 used a position angle of $27^{\circ}$ east of north and UT 2016-12-05 spectra used a position angle of $25^{\circ}$ east of north.  The sets of spectra from both nights were reduced using a processing pipeline built from standard IRAF tools, and individual exposures were median combined to remove artifacts from cosmic rays.  Wavelength calibration was done using spectra taken of Helium, Neon, and Argon lamps for the DIS spectra and Argon, Mercury, Neon, and Cadmium lamps for the DeVeny spectra on each respective night.  As part of each observing night, dome flats were used for flat fielding and standard stars from the \citet{1988ApJ...328..315M} catalog were observed for flux calibration. After reduction, the corresponding red and blue spectra were truncated in an overlapping region and stitched together.  To correct for the intermittent clouds that affected the observations on UT 2016-12-05, the observed flux was scaled such that the integrated [OIII] emission from the $4959 \AA$ and $5007 \AA$ lines is equal to that observed on UT 2011-08-07.

\subsection{HST Imaging}

Imaging in four filters with the WFC3 camera on HST was done over three orbits on UT 2016-11-09 to achieve the observing goals outlined in Section \ref{sec-disc_samp_goals}, and they are summarized in Table \ref{table-hst_obs}.  The filters were chosen to target specific features of the galaxy.  The F814W filter is a wide optical/NIR filter that samples the old stellar population and provides insight into the galaxy structure.  Its broad wavelength range allowed our deep exposures to capture moderately faint structures in the system.  The F547M filter covered the narrow line [OIII] emission from the galaxy, as well as the narrow and broad $H\beta$ lines, which provided a window into emission from the AGN and also into emission from star formation occurring within the galaxy.  UV imaging with the F225W filter was used to help localize the thermal emission of the AGN accretion disk and search for additional unobscured AGNs, if they might have been present.  Finally, imaging with the very narrow FQ750N filter (70 \AA) was used to locate the source of the kinematically shifted broad Balmer lines.  The filter fortuitously extended only over the broad $H\alpha$ line emission with no contamination from neighboring narrow line optical emission.  With a spatial resolution of $0\arcsec.04$, the WFC3 camera pixels subtend a spatial scale of 0.1 kpc at the redshift of the host galaxy.  

Two observing strategies were employed.  For the point source imaging with the F225W and FQ750N filters, the observations were split between three dithered images to improve the sampling of WFC3's PSF and allow for the removal of cosmic rays and detector artifacts.  Additional imaging and processing were needed for the long exposures with the F547M and F814W filters to minimize effects of point source saturation.  Like previous studies \citep[e.g.,][]{2006ApJ...644..133B, 2008ApJ...679.1128K, 2013ApJ...767..149B}, we circumvent the saturation issue using a series of graduated exposure times to maximize the dynamic range of our observations.  For each filter, two sets of exposures were taken with each set composed of a short, medium, and long exposure.  The nearly linear response of CCDs allows for saturated pixels to be clipped and replaced by the same pixels from a shallower image in an exposure set that have been scaled by the exposure time ratio.  Dithering was only done between sets of exposures to ensure the pointing was the same to allow for this correction.

Prior to analysis, the individual images were reprocessed from the raw images using the \emph{calwf3} package, except for the saturation correction step that we added to the processing pipeline, and \emph{AstroDrizzle} was used to remove distortion and combine the individual exposures.  Since cosmic rays remain in the combined images, they were removed using the Python implementation (\texttt{lacosmic}\footnote{https://lacosmic.readthedocs.io/}) of the L.A. Cosmic routine \citep{2001PASP..113.1420V}.  This algorithm is a variant of Laplacian edge detection that identifies cosmic rays based on the sharpness of their edges in an image.

\subsection{\emph{Chandra} Imaging and Spectra}

Our \emph{Chandra} observing program consisted of a single, 8 ksec observation of 2MASX J00423991+3017515 on UT 2017-09-25.  The ACIS-S detector was used, which has 0.5$\arcsec\:$ pixels that provide better than 1 kpc resolution within the host galaxy of 2MASX J00423991+3017515 and an energy coverage from 0.2-10 keV.  A 0.3 s frame-time was used to mitigate pile-up and a 1/8$^{th}$ subarray was used to recover observing efficiency by reducing unnecessary deadtime.  The count rate of the source in these observations was $0.34$ cts/s, so pile-up was not a major concern.  The 128 rows of the subarray covered 64$\arcsec$ on the sky and fully contained our galaxy.  The data presented here were reduced from the raw data using the \emph{CIAO} suite of analysis tools\footnote{http://cxc.harvard.edu/ciao/}.

\subsection{\emph{Swift} Spectra}

As part of the BAT all-sky survey, all three instruments on \emph{Swift} observe at each pointing.  In our X-ray spectral modeling, we include the BAT spectrum, taken from the HEASARC archives\footnote{https://heasarc.gsfc.nasa.gov}.  We also include the longest XRT observation available (obsid $00040888001$ taken on UT 2010-11-15) with an exposure time of $2632$ seconds.  The spectra were extracted following the \emph{XSelect} guidelines using the standard tools.

\subsection{JVLA Imaging}

The target was observed by the \emph{Jansky Very Large Array} (JVLA) in the K-band ($\nu_c = 22$~GHz) on UT 2020-02-27 as part of a radio imaging survey of BAT AGN \citep{2016ApJ...832..163S,2020MNRAS.492.4216S}. The science integration was 9.0 minutes in duration and was part of a 1-hour scheduling block consisting of 3 total science targets. The observing block began with X- and K-band attenuation scans, followed by flux and bandpass calibration with 3C~138. The science integration on 2MASX J00423991+3017515 was preceded by a pointing calibration scan and bracketed by gain calibration scans. After collection, the data were passed through the standard JVLA reduction pipeline at the National Radio Astronomy Observatory (NRAO) and reduced using the Common Astronomy Software Applications package \citep{2007ASPC..376..127M}. The science target was split from the parent measurement set and averaged over the 64 spectral window channels, then cleaned to a 0.03~mJy threshold with uniform weighting. The resulting image has a $1\sigma$ sensitivity level of 9.8~$\mu$Jy.  

\section{Results}
\label{sec-results}

\subsection{Optical Spectra}
\label{sec-optical_spec_results}

 \begin{figure}
\centering
  \subfigure[UT 2011-08-07 H$\gamma$ Region]{\includegraphics[width=0.4\textwidth]{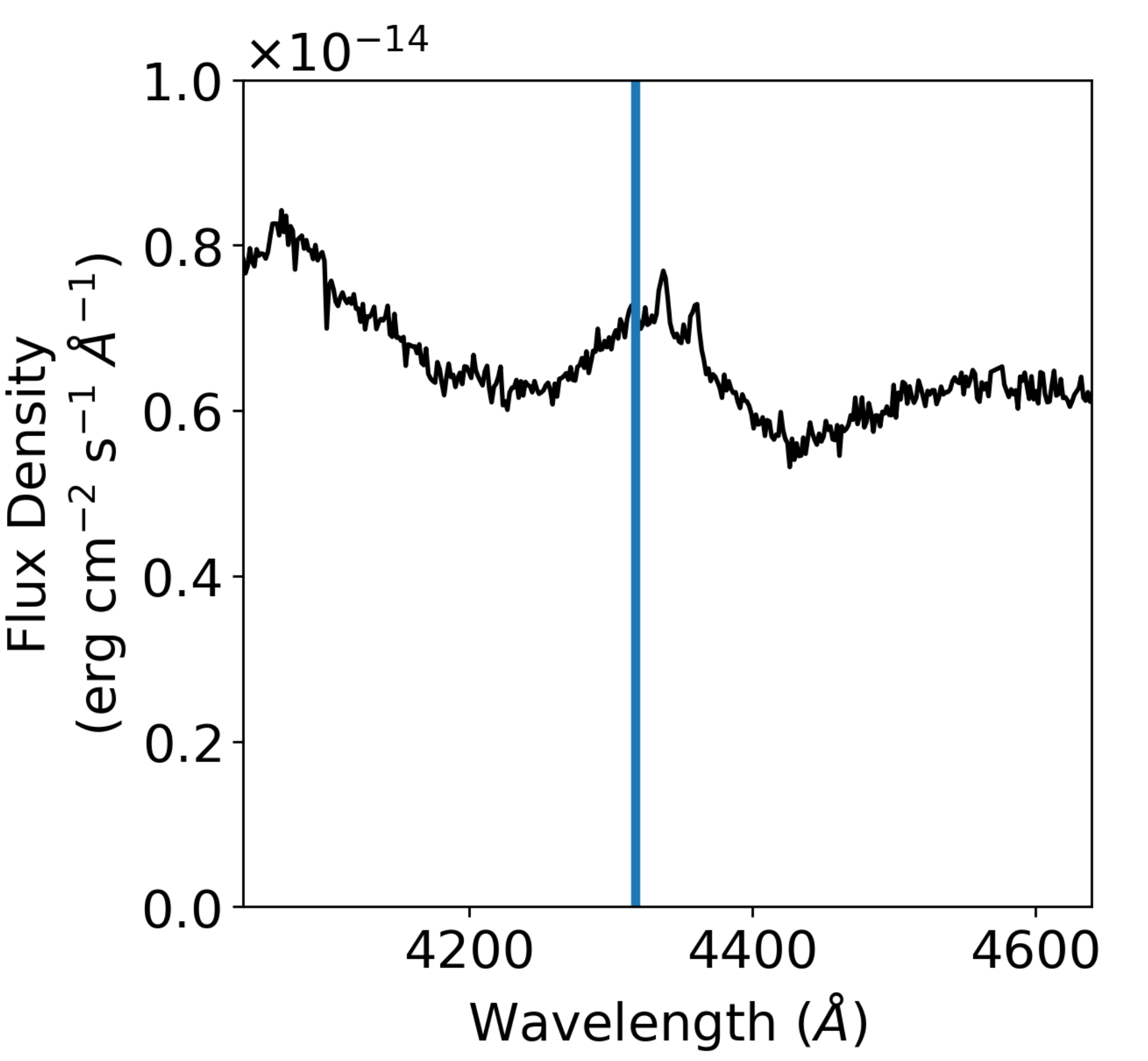} \label{fig-APO_hg}}
  \subfigure[UT 2011-08-07 H$\beta$ Region]{\includegraphics[width=0.4\textwidth]{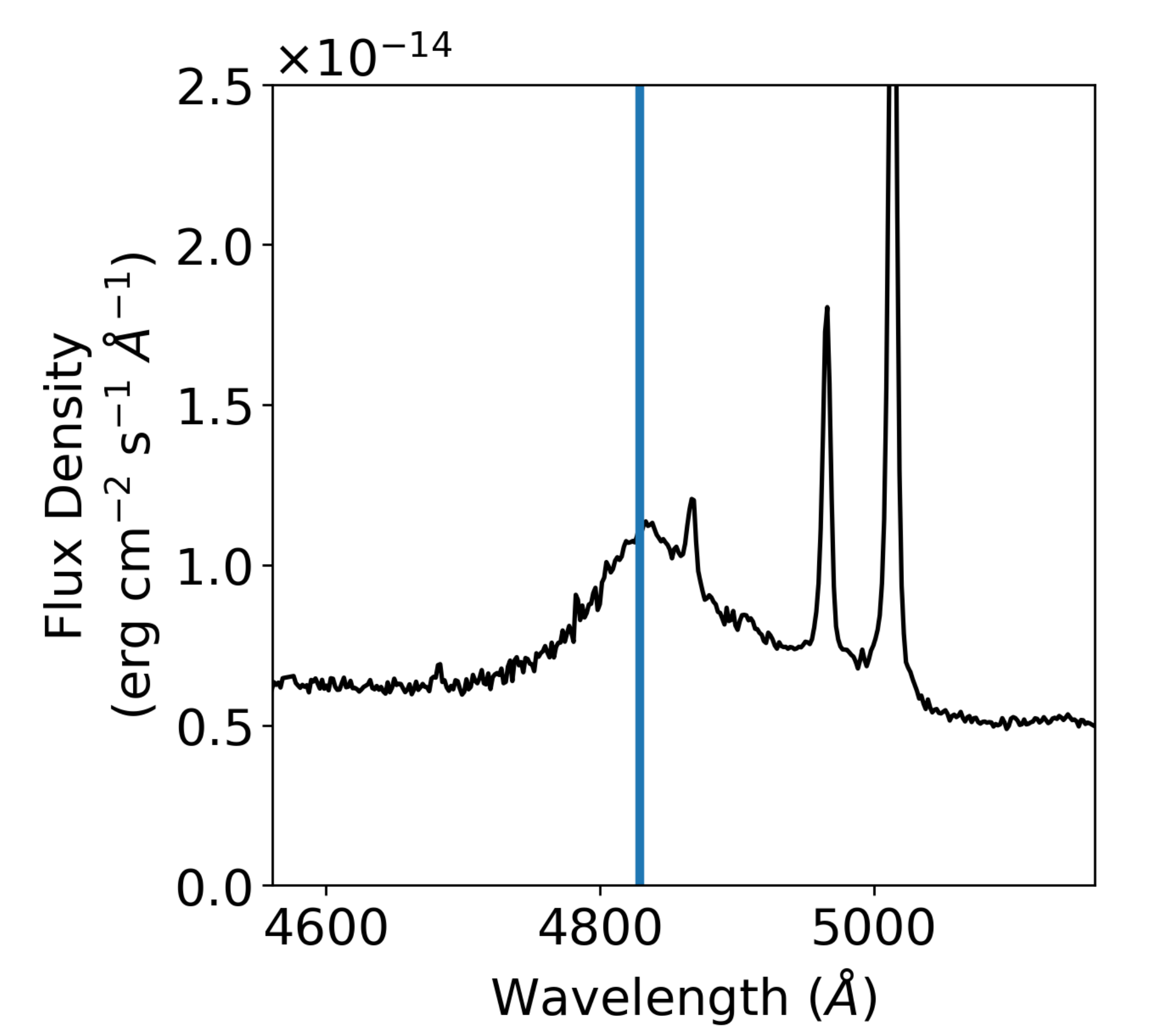} \label{fig-APO_hb}}
  \subfigure[UT 2011-08-07 H$\alpha$ Region]{\includegraphics[width=0.4\textwidth]{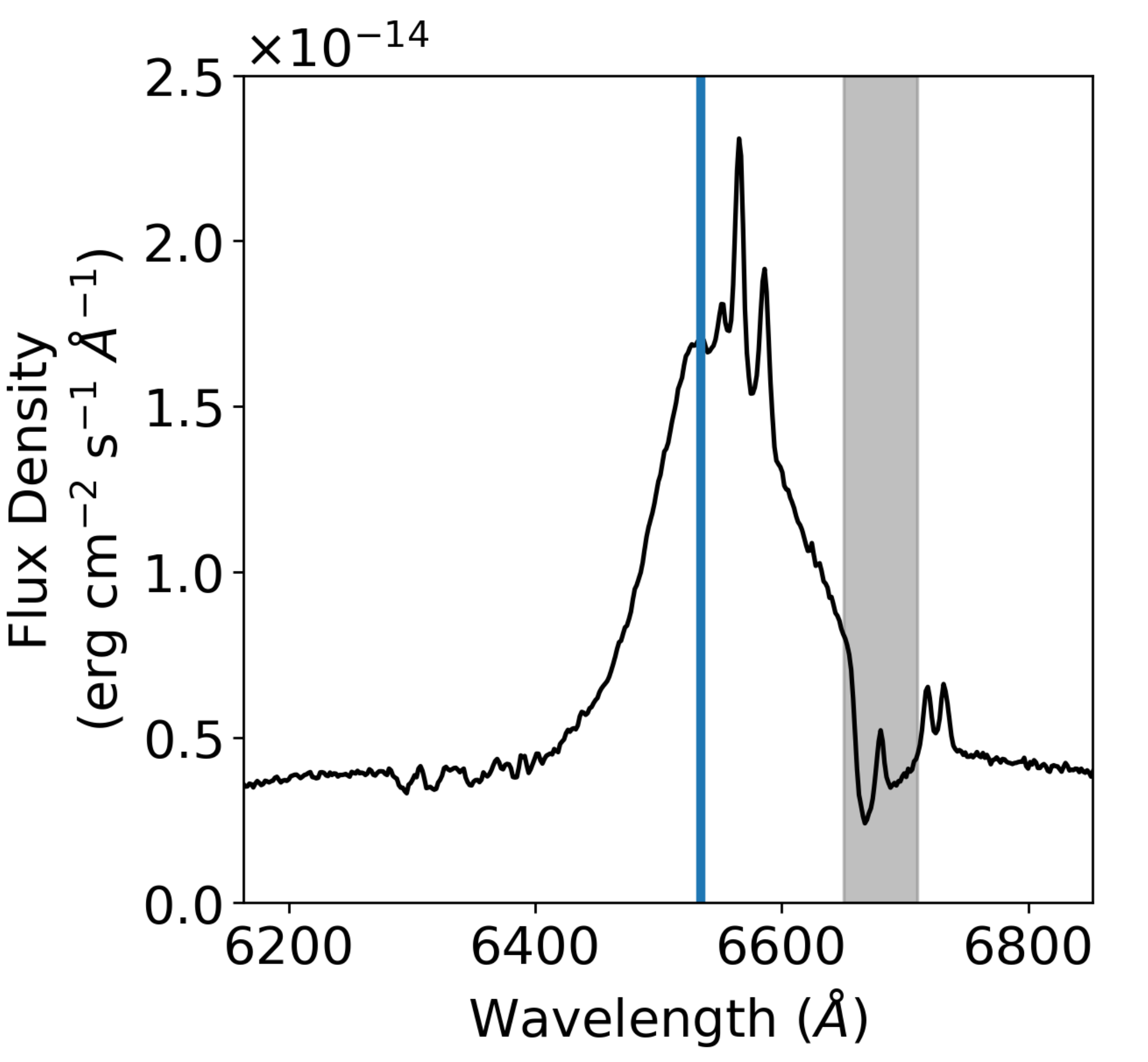} \label{fig-APO_ha}}
\caption{ The H$\gamma$ (a), H$\beta$ (b), and H$\alpha$ (c) regions of the optical spectrum taken on UT 2011-08-07 at APO.  The data is shown with black lines.  Prominent Telluric absorption contaminating the broad H$\alpha$ line is shown in grey.  The peaks of the H$\gamma$, H$\beta$, and H$\alpha$ broad lines are noticeably offset from their narrow line counterparts and their measured peaks are marked with blue vertical lines.}
\label{fig-apo_spec}
\end{figure}

\begin{figure}
\centering
  \subfigure[UT 2016-12-05 H$\beta$ Region]{\includegraphics[width=0.4\textwidth]{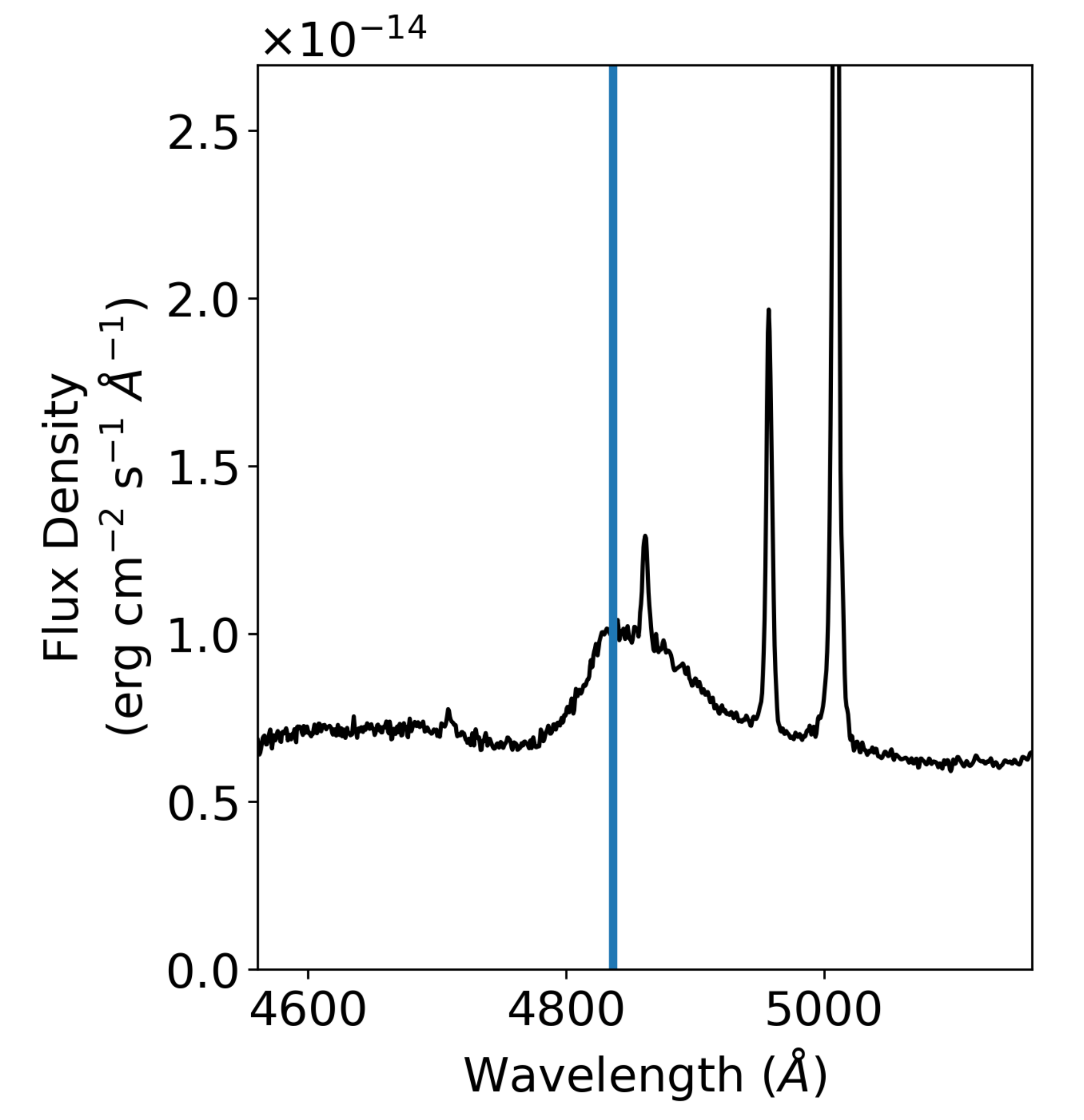} \label{fig-DCT_hb}}
  \subfigure[UT 2016-12-05 H$\alpha$ Region]{\includegraphics[width=0.4\textwidth]{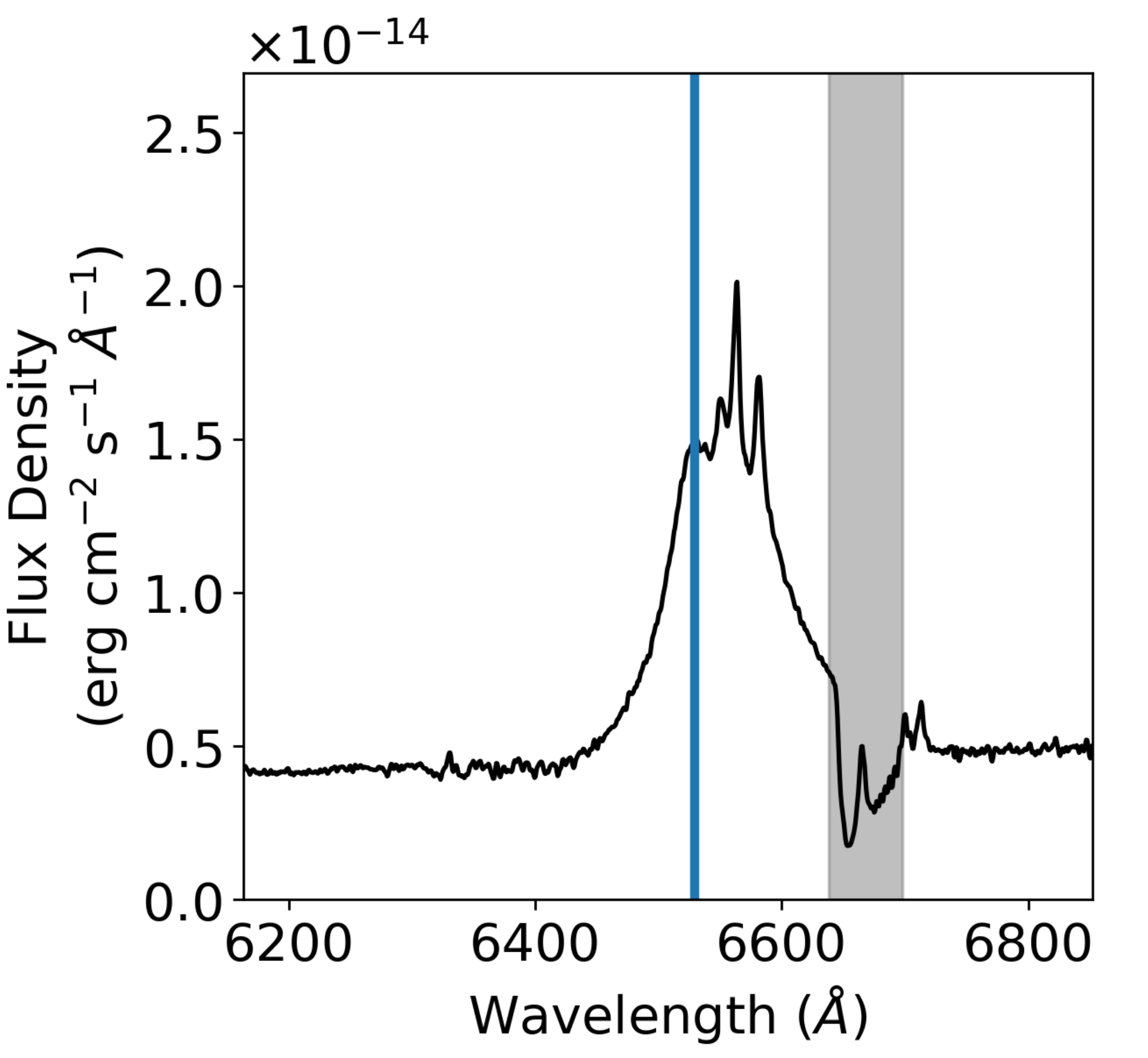} \label{fig-DCT_ha}}
\caption{The H$\beta$ (a) and H$\alpha$ (b) regions of the optical spectrum taken on UT 2016-12-05 with the LDT.  The data is shown with black lines.  Prominent Telluric absorption contaminating the broad H$\alpha$ line is shown in grey.  The peaks of the H$\beta$ and H$\alpha$ broad lines are noticeably offset from their narrow line counterparts and their measured peaks are marked with blue vertical lines.}
\label{fig-dct_spec}
\end{figure}

The optical spectra acquired with APO on UT 2011-08-07 and LDT on UT 2016-12-05 are shown in Figures \ref{fig-apo_spec} \& \ref{fig-dct_spec}, respectively.  Both spectra show the characteristic elements of a Seyfert 1 type galaxy: a single set of narrow emission lines at the same redshift as the NaII D and Mg absorption from the host galaxy, and broadened Balmer emission lines.  However, the peaks of the broad Balmer emission lines are significantly blue shifted from the corresponding peaks of the narrow Balmer emission.  To measure the shift, a Savitzky-Golay filter that smoothed over an 11 pixel window with a second degree polynomial was first applied to the spectra to remove any artificial peaks from stochastic fluctuations due to noise.  Then, the significant local maxima were found by thresholding for prominences greater than $1\times10^{-16}$ erg s$^{-1}$ cm$^{-2}$ $\AA^{-1}$ with widths larger than 5 pixels.  Once identified, the differences in the peaks between the broad lines and narrow lines were calculated.  In the galaxy's rest frame, on UT 2011-08-07 the peaks of the broadened H$\gamma$, H$\beta$, and H$\alpha$ lines were blue shifted by 22.2 $\AA$ ($-1531$ km/s), 31.9 $\AA$ ($-1967$ km/s), and 29.0 $\AA$ ($-1324$ km/s) for an average shift of $-1600\pm300$ km/s.  On UT 2016-12-05 the peaks of the broadened H$\beta$ and H$\alpha$ lines on were blue shifted by 24.5 $\AA$ ($-1512$ km/s) and 31.4 $\AA$ ($-1437$ km/s) for an average shift of $-1480\pm180$ km/s.  These measured peaks are marked in Figures \ref{fig-apo_spec} \& \ref{fig-dct_spec}.  Errors were determined using an iterative Monte Carlo method that generated 500 fake data sets drawn from the measurement uncertainties and repeated the offset measurements.  There is no statistical difference in these measurements.

Some variability in the line shapes is noticeable between the two spectra.  The most obvious change occurs in the broad H$\beta$ line where the blueward wing shows a decrease and marked steepening.  Additionally, the broad H$\alpha$ line displays a narrowing.  Nonetheless, a large, statistically significant offset of $> 1300$ km s$^{-1}$ clearly persists and is consistent across the 5.5 yr time baseline between these observations.  Given the suboptimal conditions in which the UT 2016-12-05 spectra were taken, as mentioned in Section \ref{sec-disc_samp_goals}, Appendix \ref{sec-ldt_systematics} provides a brief discussion of our efforts to understand systematic uncertainties related to the data acquisition and reduction that might underlie the apparent differences in the line shapes.

\begin{figure}
\centering
  \subfigure[UT 2011-08-07 H$\beta$ Region Fit]{\includegraphics[width=0.4\textwidth]{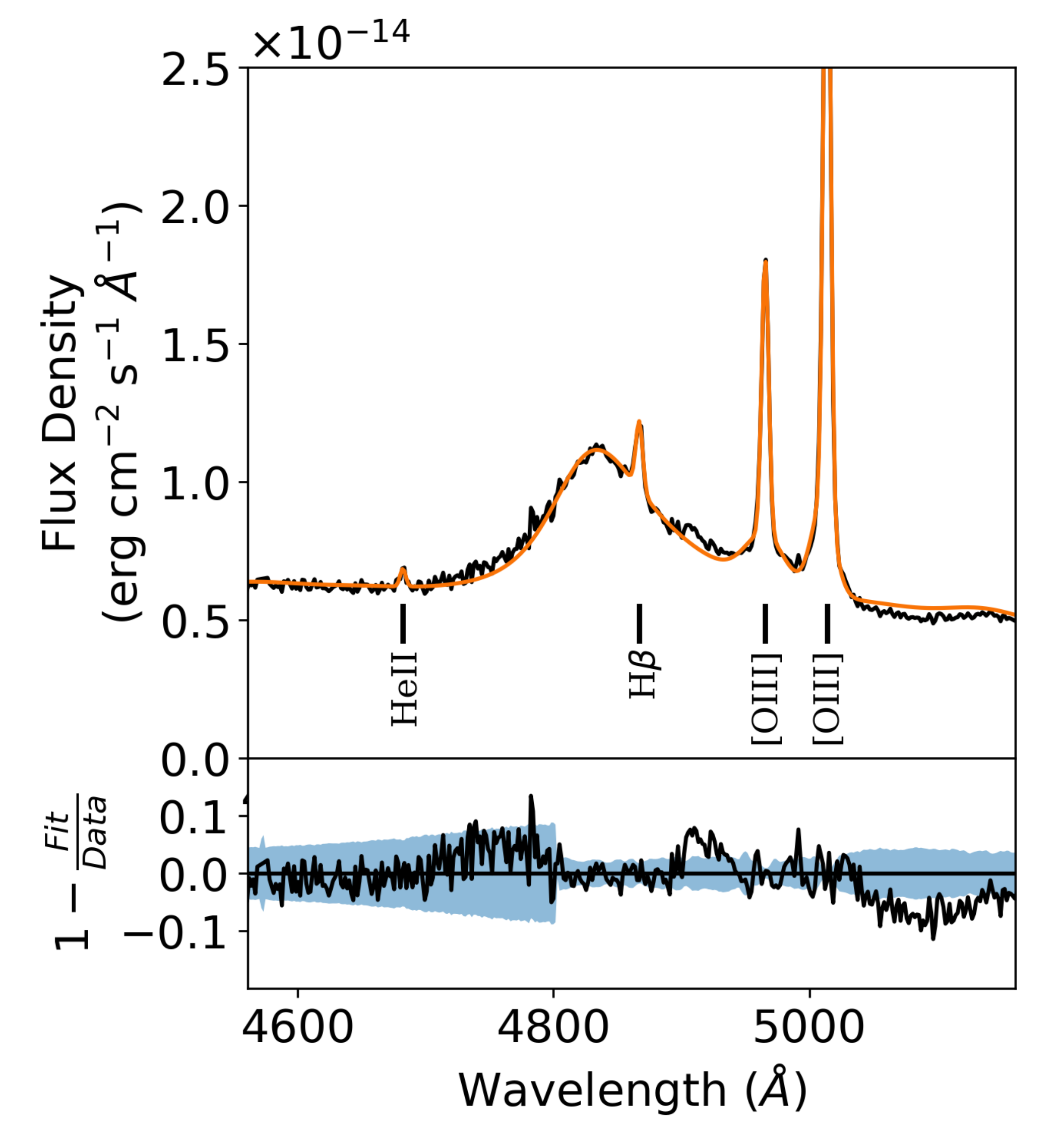} \label{fig-APO_hb_fit}}
  \subfigure[UT 2016-12-05 H$\beta$ Region Fit]{\includegraphics[width=0.4\textwidth]{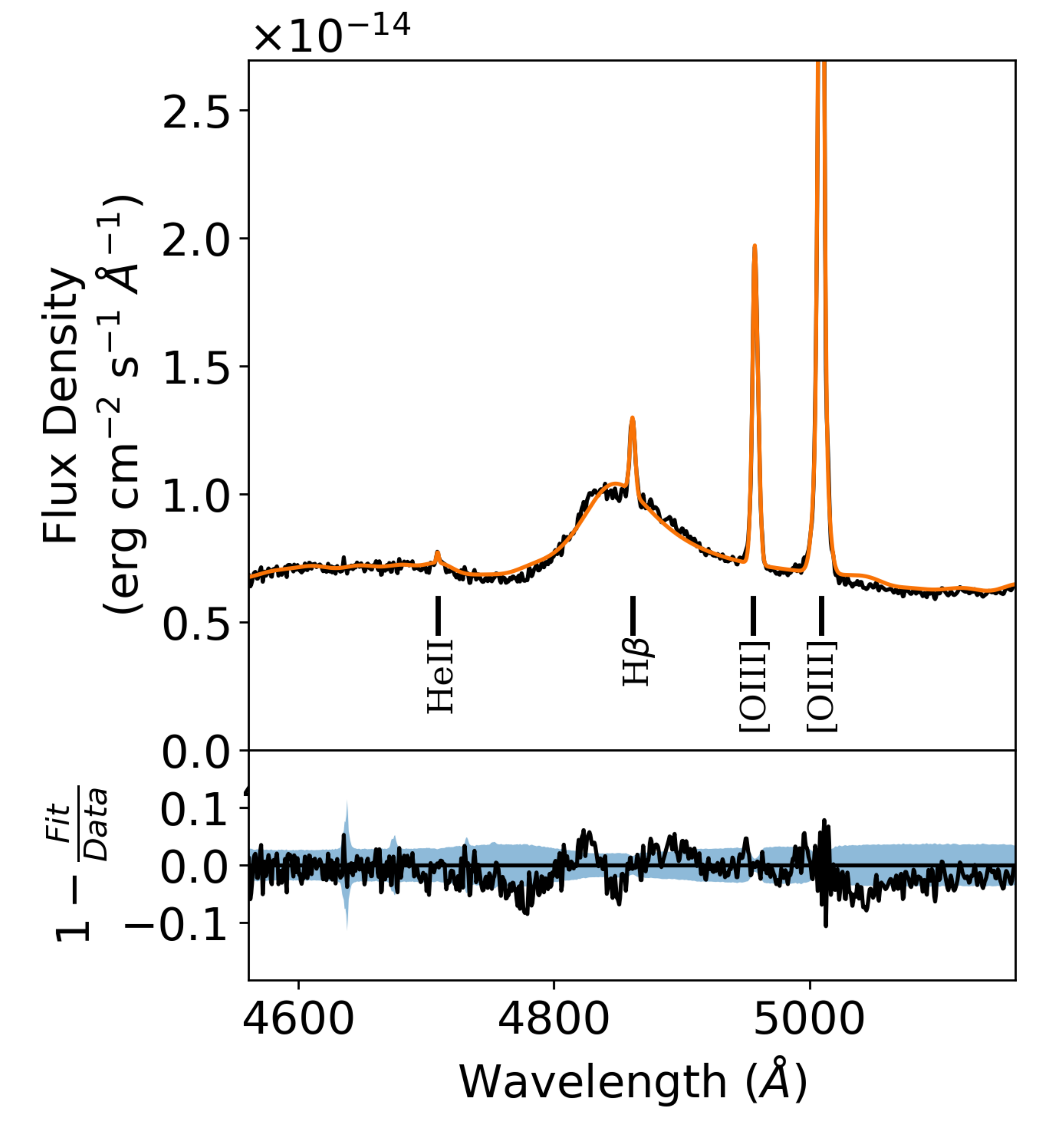} \label{fig-DCT_hb_fit}}
\caption{The fitted H$\beta$ regions of the optical spectra from UT 2011-08-07 (a) and UT 2016-12-05 (b).  The data is shown with black lines and the spectral fits used for SMBH mass estimation are shown with as orange lines.  Below each spectrum the fit residuals are shown as the percentage of unexplained flux with the percentage uncertainty at each wavelength bin for the two spectra plotted as blue bands.  Note the sharp change in uncertainty band for the UT 2011-08-07 spectrum is an artifact of stitching the red and blue spectra.}
\label{fig-hb_fit}
\end{figure}

Using the full-width half-maximum (FWHM) of the broad H$\beta$ lines and the continuum levels at $5100$ $\AA$ in the rest frame of the galaxy, single epoch black hole mass estimates can be made.  The empirical mass scaling relationship established by \citet{2006ApJ...641..689V}, \begin{equation}
\text{log M$_{BH}$} = \text{log}\Bigg\{\Bigg[ \frac{\text{FWHM(H$\beta$)}}{1000 \text{ km s}^{-1}}\Bigg]^2\Bigg[\frac{\lambda \text{F}_\lambda(5100)}{10^{44} \text{erg s$^{-1}$}} \Bigg]^{0.5}\Bigg\} + 6.91,
 \end{equation} relates the BLR size and luminosity to SMBH mass, and was calibrated using emission-line reverberation measurements.  To fit the H$\beta$ lines in each spectrum, the line was isolated by fitting a power law continuum, $F\propto\nu^{\alpha}$, which takes the form, \begin{equation}
F = C \lambda^{2-\alpha},
\end{equation}
where $F$ is the flux, $C$ is the normalization, and $\alpha$ is the spectral index, and subtracting it.  Then, the He II, broad H$\beta$, narrow H$\beta$ and [OIII] emission lines were fit with Gaussian profiles.  The broad wings at the base of the [OIII] emission lines are fit with a second Gaussian.  The broad Balmer lines have an asymmetry that required skewness to also be fit, so this additional parameter was included, and a positive (blueward) skew was preferred for each broad line.  The fits to the regions are shown in Figure \ref{fig-hb_fit}.
 
In the UT 2011-08-07 spectrum we measure the H$\beta$ FWHM to be $107.0$ $\AA$ ($6640$ km s$^{-1}$) and the $5100$ $\AA$ continuum to be $5.1\times 10^{-15}$ erg s$^{-1}$ cm$^{-2}$ $\AA^{-1}$ ($\lambda$L$_{\lambda}$ = $1.38\times10^{45}$ erg s$^{-1}$). In the UT 2016-12-05 spectrum we measure the H$\beta$ FWHM to be $80.2$ $\AA$ ($4974$ km s$^{-1}$) and the $5100$ $\AA$ continuum to be $6.0\times 10^{-15}$ erg s$^{-1}$ cm$^{-2}$ $\AA^{-1}$ ($\lambda$L$_{\lambda}$ = $1.61\times10^{44}$ erg s$^{-1}$).  This yields black hole mass estimates from the UT 2011-08-07 and UT 2016-12-05 spectra of M$_{BH}$ = $10^{9.1\pm0.3}$ and M$_{BH}$ = $10^{8.5\pm0.5}$, respectively.

\begin{figure*}
\centering
  \subfigure[F814W and F225W Images ]{\includegraphics[width=\textwidth]{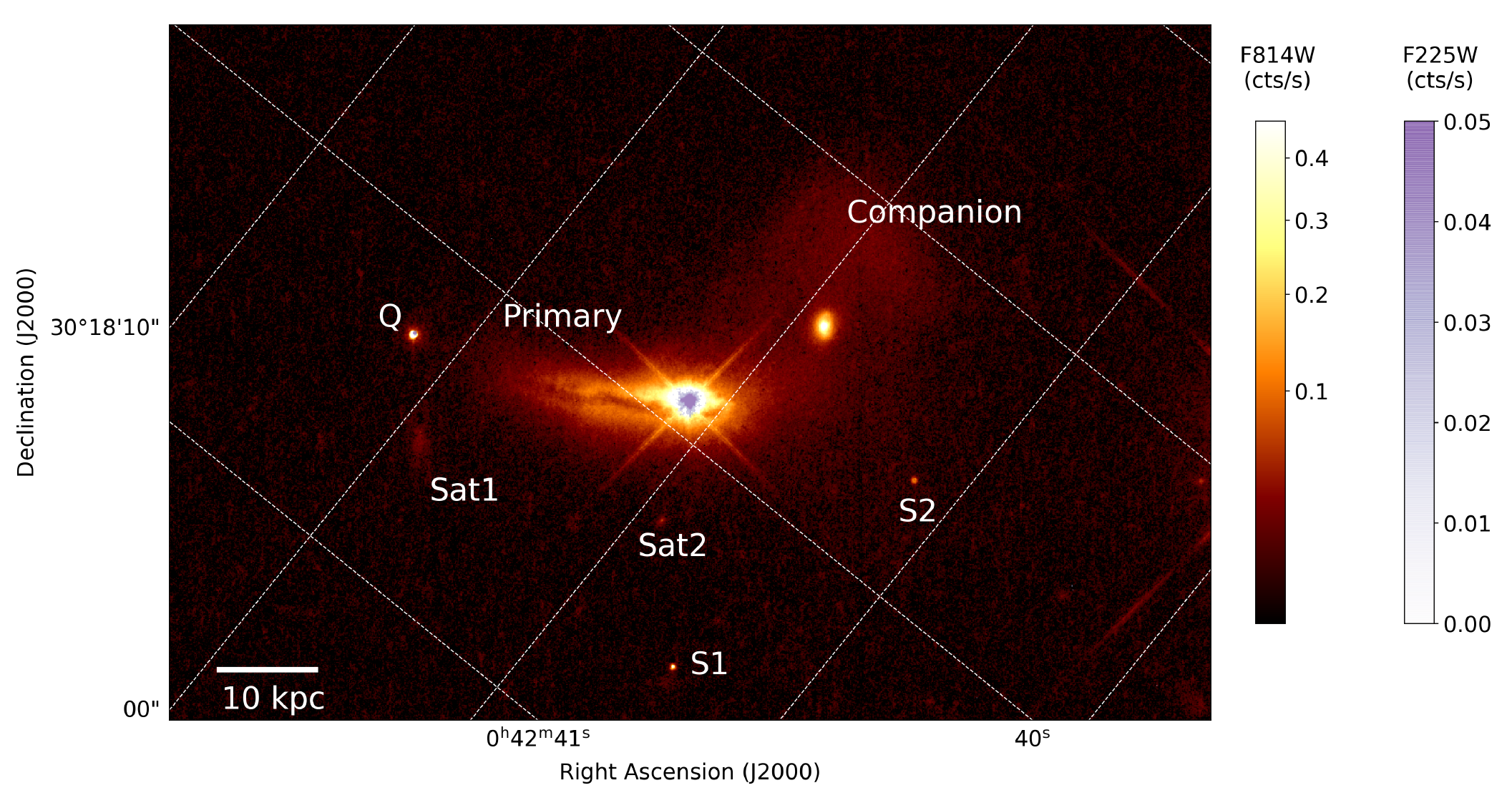} \label{fig-F814W_F225W}}
  \subfigure[F547M and FQ750N Images]{\includegraphics[width=\textwidth]{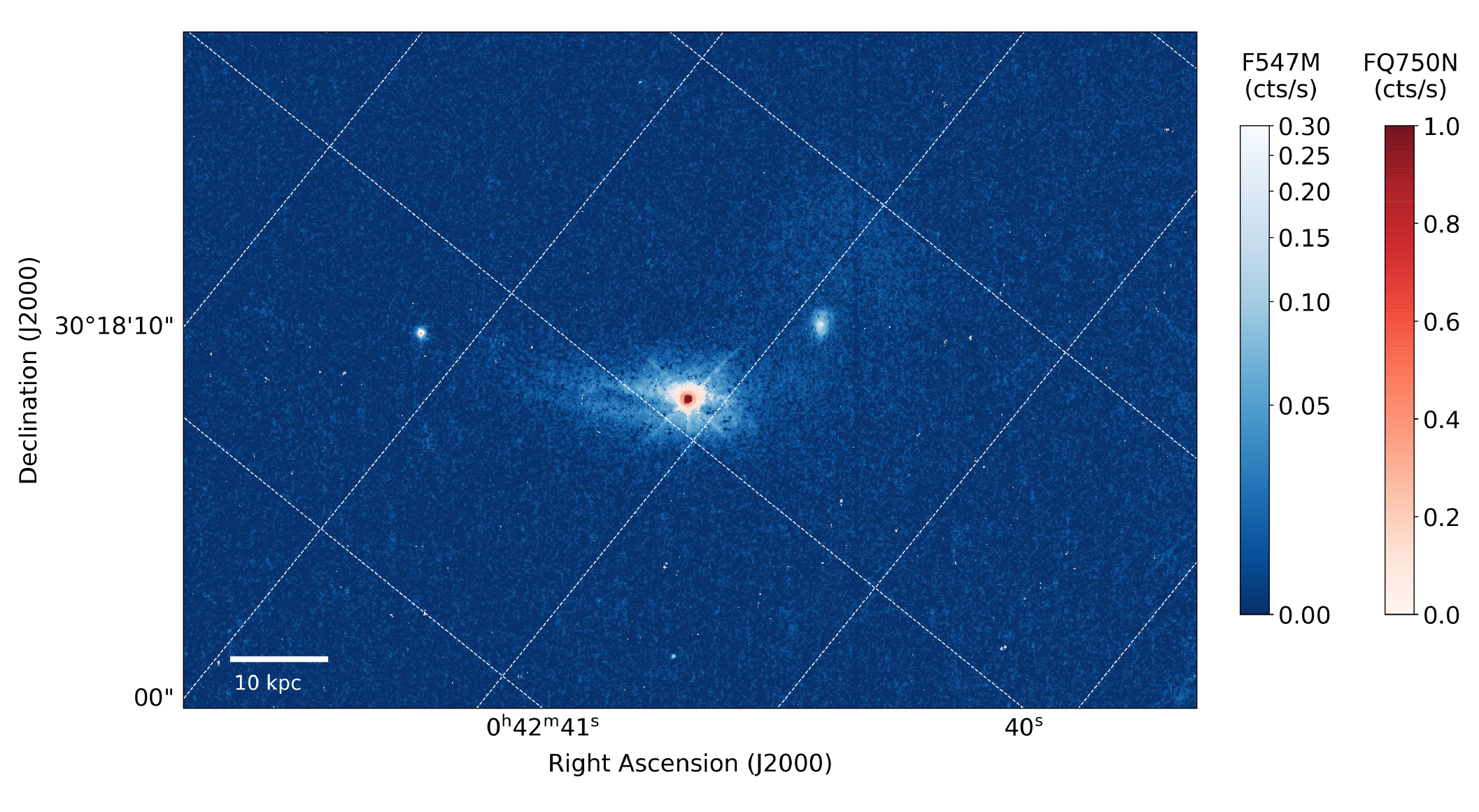} \label{fig-F547M_FQ750N}}
\caption{{HST images taken in the F814W and F225W filters (a) and in the F547M and FQ750N filters (b).  Due to the high count rates and need for better contrast to accentuate low surface brightness features, the count rates of the F814W and F547M filters are shown on a logarithmic scale.  The count rates of the F225W and FQ750N filters are shown on a linear scale.}}
\label{fig-hubble_imaging}
\end{figure*}

\subsection{HST Imaging}
\label{sec-hubble_results}

The HST imaging of 2MASX J00423991+3017515 is shown in Figure \ref{fig-hubble_imaging}.  The F814W image reveals 7 objects in the field.  Central to the image is the primary galaxy.  It contains a single bright AGN (located at the intersection of the single set of diffraction spikes) and is interacting with a faint companion.  A dark dust lane bisects the primary galaxy, indicating we are viewing a nearly edge-on disk.  It shows a slight perturbation, but it is mostly undisturbed, which suggests the interaction we are observing is weak and in the early stages of the merger.  There could be alternate interpretations, though, which we provide in Section \ref{sec-discussion}.  The smaller companion galaxy has diffuse stellar morphological features, which are telling of an ongoing interaction between the two. Another compact, marginally-resolved source, \emph{Sat1}, is located on the opposite side of the primary along with a second satellite galaxy, \emph{Sat2}.  \emph{S1} and \emph{S2} appear to be foreground stars, and \emph{Q} is a background quasar.  The primary galaxy and its companion will be the main subjects of the analysis.

The F547M filter, covering [OIII] and H$\beta$, shows that most of the emission comes from the AGN itself and the region immediately surrounding it, but there is also a significant extended flux originating from the AGN hosting galaxy.  The nuclear region of the companion is also bright, along with the background quasar, \emph{Q}.

The F225W filter only shows a single point source that is coincident with the AGN and trace emission from \emph{Q}. This strongly indicates that there is only one unobscured AGN in the system.  The FQ750N filter, which captures only the offset broad H$\alpha$ emission, also shows only a point source that is co-located with the AGN.  This makes the interpretation of the optical spectra in Section \ref{sec-optical_spec_results} easier, since this point source is solely responsible for AGN-related optical emission.  \emph{Q} weakly appears in the FQ750N filter.

The GALFIT package \citep{2002AJ....124..266P, 2010AJ....139.2097P} was used to quantify the structure and morphology of the two galaxies by fitting analytic functions to the 2-D light profiles of the system.  GALFIT uses the point spread function (PSF) of the telescope to model unresolved point sources like an AGN and to convolve with the model components.  Ideally, the PSF is measured empirically from bright stars within the field-of-view of the telescope during the observation, but no suitable stars were present.  The only two bright stars in our images were so bright that the core of the PSF was saturated, even during our short 30s exposure.  Instead, we simulated the PSF using TinyTim \citep{1995ASPC...77..349K}.

Except for the AGN, the other components are modeled as S\'ersic profiles \citep{1963BAAA....6...41S} of the form, \begin{equation}
\Sigma(r) = \Sigma_e \textrm{exp}\bigg[ -\kappa \bigg( \bigg(\frac{r}{r_e}\bigg)^{\frac{1}{n}}-1\bigg)\bigg)\bigg].
\end{equation}  The function depends on the half-light radius ($r_e$), the surface brightness at that radius ($\Sigma_e$), and the index ($n$) which controls the degree of curvature of the profile.  The constant $\kappa$ is tied to $n$ and approximately equal to $2n-1/3$.  By varying $n$, the S\'ersic profile provides the flexibility to describe a wide range of typical galactic components including exponential disks ($n\approx1$) and classical bulges ($n>2$).  Lopsidedness and other warping of the profiles were allowed for in our fitting by Fourier mode modifications to the S\'ersic profiles, but only two components needed them implemented.  The dust lane was masked so that it would not contaminate our model.

\begin{figure*}
\centering
 \includegraphics[width=\textwidth]{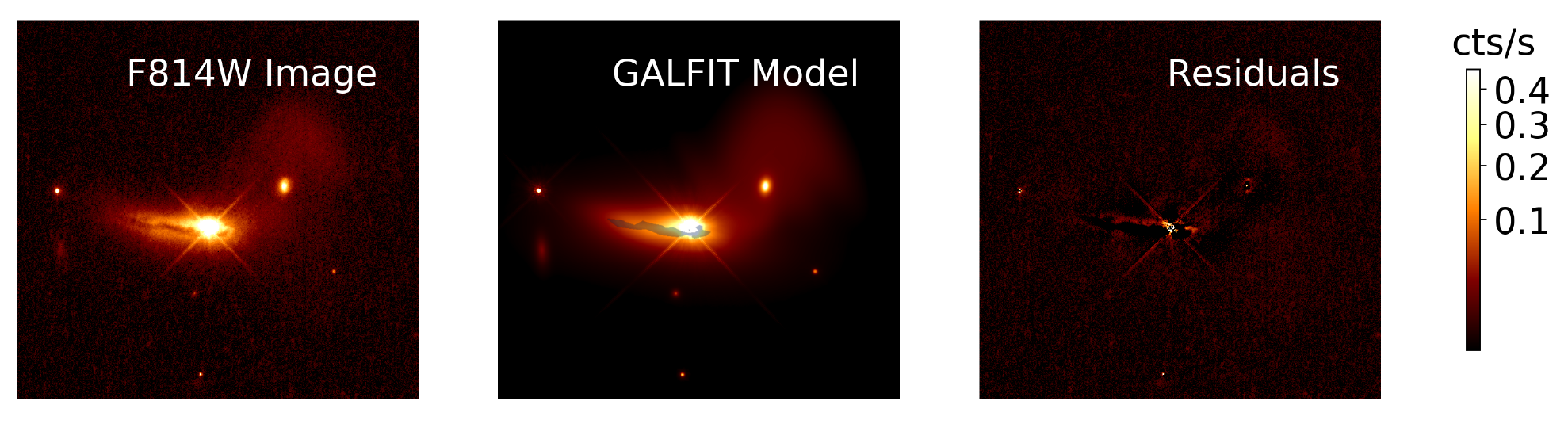}
\caption{The F814W filter HST image of 2MASX J00423991+3017515 (left), the summed components of the best fitting GALFIT model (center), and the residuals unexplained by the fit (right), shown on the same scale as the F814W image in Figure \ref{fig-hubble_imaging}. The dust mask is overlaid in blue.}
\label{fig-galfit_plot}
\end{figure*}

\begin{table*}
	\centering
	\caption{GALFIT Model Parameters for Primary Galaxy and Interacting Companion}
	\label{table-galfit}
	\begin{threeparttable}
	\begin{tabular}{c|cccccc} 
		\hline
    Parameter & Component 1  & Component 2 & Component 3 & Component 4 & Component 5 & Component 6 \\
		\hline
    Type & PSF & S\'ersic  & S\'ersic & S\'ersic & S\'ersic & S\'ersic \\
    Position x\tnote{1} 					& $1265.71$ 		& $1263.90$ 		& $1227.31$ 	& $1263.90$		& $1408.41$ 	& $1408.49$\\
    Position y\tnote{2}				& $1304.70$ 		& $1302.33$ 		& $1304.78$ 	& $1302.33$		& $1384.15$ 	& $1384.15$\\
    Integrated Magnitude\tnote{3}		& $17.81$ 		& $18.95$ 		& $19.58$ 	& $18.86$			& $21.24$ 	& $19.68$\\
    Effective Radius\tnote{4} 			& -- 				& $46.46$ 		& $75.48$ 	& $1.37$			& $8.74$ 		& $169.40$\\
    S\'ersic Index 				& -- 				& $1.71$ 			& $1.15$ 		& $3.45$			& $1.54$ 		& $0.24$\\ 
    Axis Ratio\tnote{5} 				& -- 				& $0.42$ 			& $0.35$ 		& $0.53$			& $0.67$ 		& $0.40$\\
    Position Angle\tnote{6} 				& -- 				& $76.67$ 		& $-88.42$ 	& $87.45$			& $-8.58$ 		& $-25.69$\\
    Mode 1 (Amp., $\phi$) 		& -- 				& --		 		& --		 	& $-0.80, 61.28$	& -- 			& $0.31, 72.89$\\
    Mode 3 (Amp., $\phi$) 		& -- 				& --			 	& --		 	& --				& -- 			& $0.13, -57.86$\\
    Mode 4 (Amp., $\phi$) 		& -- 				& --			 	& --		 	& $0.09, -32.71$	& -- 			& $0.07, -30.06$\\
    Spiral Outer Radius\tnote{7}	& -- 				& --			 	& --		 	& --				& -- 			& $8.03$\\
    Cum. Coordinate Rotation\tnote{8}	 	& -- 				& --			 	& --		 	& --				& -- 			& $-48.65$\\
    Asymptotic Power Law 		& -- 				& --			 	& --		 	& --				& -- 			& $2.42$\\
    Inclination to Line-of-sight\tnote{9} 		& -- 				& --			 	& --		 	& --				& -- 			& $62.19$\\
    Association\tnote{10} & AGN & Primary & Primary & Primary & Companion & Companion \\

		\hline
	\end{tabular}
	\begin{tablenotes}
	    \item[1] x-position of component center in image pixels.
	    \item[2] y-position of component center in image pixels.
	    \item[3] Total brightness of component in magnitudes.
	    \item[4] Effective radius of S\'ersic profile, measured in pixels.
	    \item[5] Axis ratio (b/a).
	    \item[6] Position angle of component in degrees on the image (up=0, left=90).  The image itself was taken with a position angle of $-51^{\circ}$ east of north.
	    \item[7] Radius beyond which spiral behaves like a pure power law, measured in pixels.
	    \item[8] Total rotation to outer radius in degrees.
	    \item[9] Inclination to line of sight in degrees (face-on=0, edge-on=90).
	    \item[10] Physical object associated with the model component, either the observed AGN, the AGN hosting ``Primary" galaxy, or the ``Companion" galaxy.

	\end{tablenotes}
	\end{threeparttable}
\end{table*}

The best fitting GALFIT model to the F814W image has $\chi^2/DOF=1134042.28/548294 = 2.07$, shown in Figure \ref{fig-galfit_plot}.  The model of the AGN hosting galaxy and the interacting companion is built with a PSF for the AGN, three S\'ersic profiles for the AGN hosting galaxy, and two S\'ersic profiles for the companion.  The parameters for these components are presented in Table \ref{table-galfit}.  In the primary galaxy, the body is best described with a standard exponential disk ($n\approx1$).  The PSF component (Component 1) has its center $38.4\pm0.3$ pixels, or $1\arcsec.52\pm0.01$, away from the center of the disk component (Component 3), a distance of $3.78\pm0.02$ kpc at the redshift of the system.  The other two S\'ersic profiles in the fit of the primary galaxy are centered around the PSF and correspond to a compact region (Component 4, $r_e = 1.37$ pixels) and another, moderately extended region (Component 2, $r_e=46.46$).  Both components have eccentricities of $\approx0.4$.  Interpreting these components is difficult because of their shape and position.  Component 4 has a high S\'ersic index like a bulge, but it is physically small.  Component 2, on the other hand, is large like a bulge, but more ``disk-like" since it has a lower S\'ersic index than the typical bulge.  Furthermore, these components appear to be offset from the primary disk while also being coincident and inline with the region producing the bulk of the line emission in the F547M filter, presumably a region with an abundance of star formation.  Lacking dynamical information on the system, it is reasonable to attribute these features to a stretched bulge where the disk material is disrupted from the encounter with the companion or from a lopsided enhancement in star formation.

The companion is best fit as a galaxy with a bulge and is centered 198 pixels from the center of the primary galaxy, or $7\arcsec.8$.  The body of the companion is elongated, inclined to our line of sight, and requires modification with Fourier modes and a logarithmic spiral mode to account for asymmetries in its geometry, presumably resulting from the close passage to the AGN host.

In the GALFIT model \emph{Q}, \emph{S1}, and \emph{S2} are modeled as point sources.  The other faint satellites, \emph{Sat1} and \emph{Sat2}, are modeled as S\'ersic profiles.  {\emph{Sat1} has an integrated magnitude of $23.87$ and its center is $248$ pixels, or $10$ kpc, from the center of the exponential disk component the AGN hosting galaxy.  The light profile has $n=0.83$, $r_e=18.0$, and an axis ratio of $0.42$.  \emph{Sat2} has an integrated magnitude of $25.37$ and its center is $127$ pixels, or $5$ kpc, from the center of the AGN hosting galaxy's main disk component.}  The light profile has $n=1.35$, $r_e=3.1$, and an axis ratio of $1.0$.

Mapping the residuals from the fit shows the model does well on the bulk of the galaxy's emission, but there is additional small scale structure around the nucleus and the edge of the dust lane in the plane of the galaxy that is not fully captured.  The likely source of these residuals is star formation in the plane of the galaxy that the dust lane has mostly obscured.

\begin{figure*}
\centering
 \includegraphics[width=\textwidth]{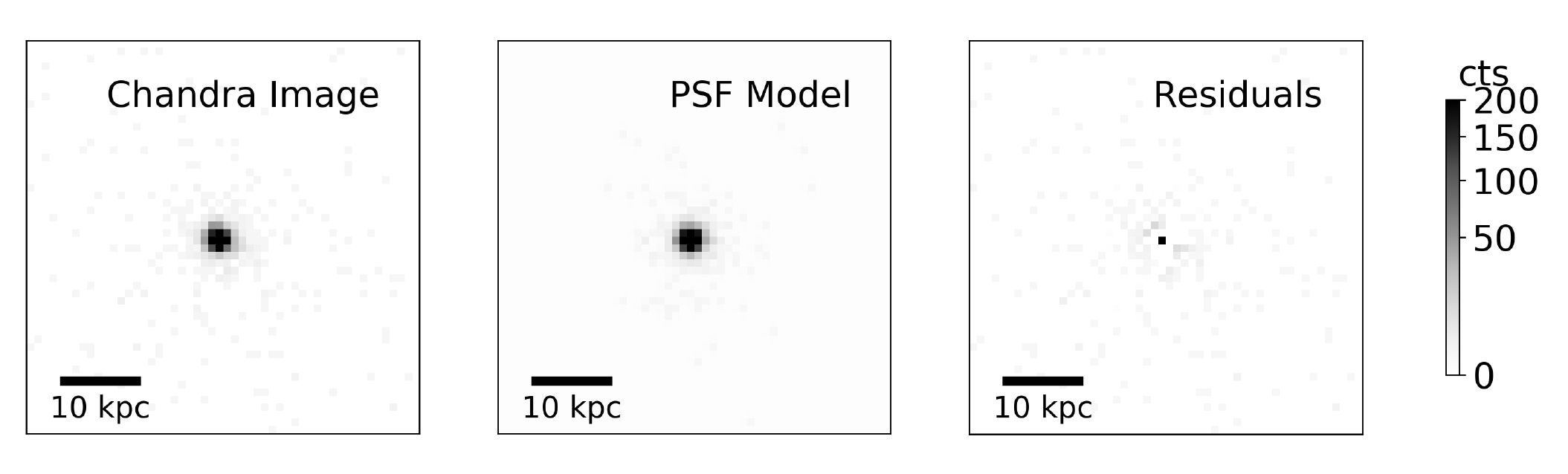}
\caption{The X-ray image of 2MASX J00423991+3017515 taken with the ACIS instrument on \emph{Chandra} (left), the fit PSF model (center), and the residuals unexplained by the fit (right), all shown on the same scale. There are a few residual counts observed from inaccuracies in the PSF at the peak of the image and symmetrically distributed in the extended wings, but no clear indication of structure or secondary sources of emission in the image.}
\label{fig-chandra_fit}
\end{figure*}

\subsection{\emph{Chandra} Imaging}

\begin{figure*}
\centering
 \includegraphics[width=0.75\textwidth]{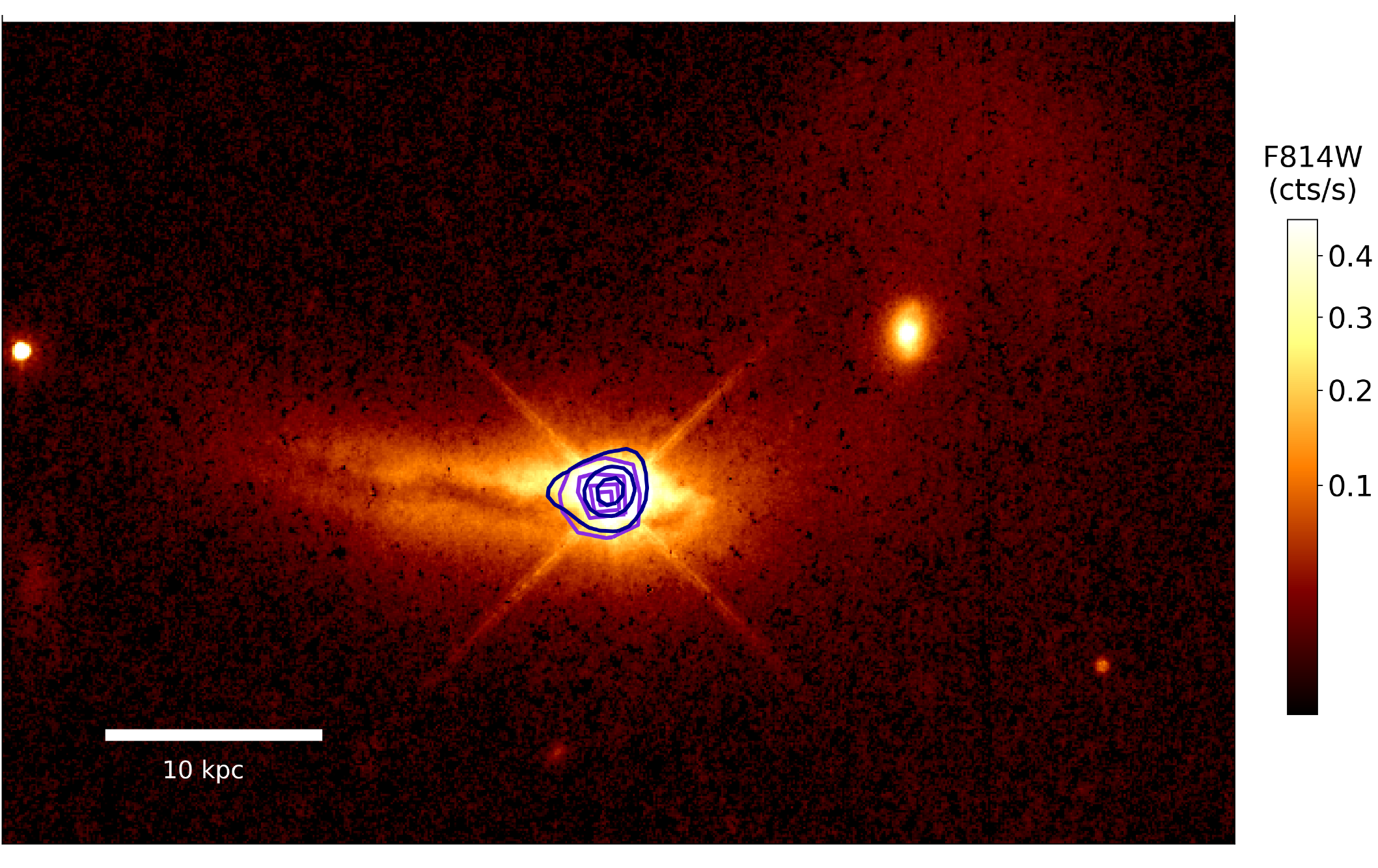}
\caption{The F814W filter HST image shown on the same scale as the F814W image in Figure \ref{fig-hubble_imaging} with contours of the \emph{Chandra} X-ray image set at $100$, $250$, $500$, and $800$ counts overlaid in purple, and contours of the JVLA radio image set at $9\times10{-5}$, $3\times10^{-4}$, and $5\times10^{-4}$ Jy/beam overlaid in blue.  The only clear source of X-ray and radio emission is the bright AGN and no secondary emitters are found, and particularly none seen around the measured center of the main disk component in the galaxy's GALFIT model.}
\label{fig-chandra_hubble_image}
\end{figure*}

The \emph{Chandra} ACIS-S image of 2MASX J00423991+3017515 was constructed from the observed events and is shown in Figure \ref{fig-chandra_fit}.  In this image, only a single compact source is detected.  To fit the two-dimensional distribution of the X-ray emission, we used Sherpa \citep{2001SPIE.4477...76F} from the Chandra Interactive Analysis of Observations (CIAO) package \citep{2006SPIE.6270E..1VF}.  A PSF model was created by performing a detailed ray-trace simulation using ChaRT\footnote{http://cxc.harvard.edu/chart/} and then MARX\footnote{http://space.mit.edu/CXC/MARX/} was used to create the final PSF.  This was then fit to the point source with the addition of a two-dimensional Gaussian component to allow for a slight blurring to account for differences between the simulated PSF and the empirical PSF.  The best fit to the image is a PSF+Gaussian model with a pixel center at x=$4062.72$ and y=$4096.87$ with a Gaussian FWHM of 0.85 pixels.  This model has $\chi^2/DOF$ = $1403.8/2699$ = $0.52$, and is shown in Figure \ref{fig-chandra_fit}. 

The residuals show no indication that a second emitter is present in the host galaxy or the larger interacting system.  There is a slight excess of residual emission on the outskirts of the PSF profile, but it is symmetric and likely the result of deviations of the PSF model derived through ray tracing from the actual PSF.  The signature of a second AGN in the residual image would be a concentration of excess emission, and there is no evidence of such feature.  The caveat to this conclusion, of course, is that an obscured secondary AGN could be detected with much deeper X-ray observation or observation at higher energies where the obscuration effects are lessened.  There could also be another SMBH that is not accreting at all.  However, with this data and the fit performed, the presence of any additional AGNs in the system is overwhelmingly disfavored.

In Figure \ref{fig-chandra_hubble_image}, the contours of the X-ray emission are overlaid on the HST F814W image.  The emission is associated with the bright, offset AGN, indicating the optical AGN is also the lone X-ray emitter.

\subsection{X-ray Spectra}

\begin{figure}
\centering
  \subfigure[Swift XRT \& BAT Spectra]{\includegraphics[width=0.5\textwidth]{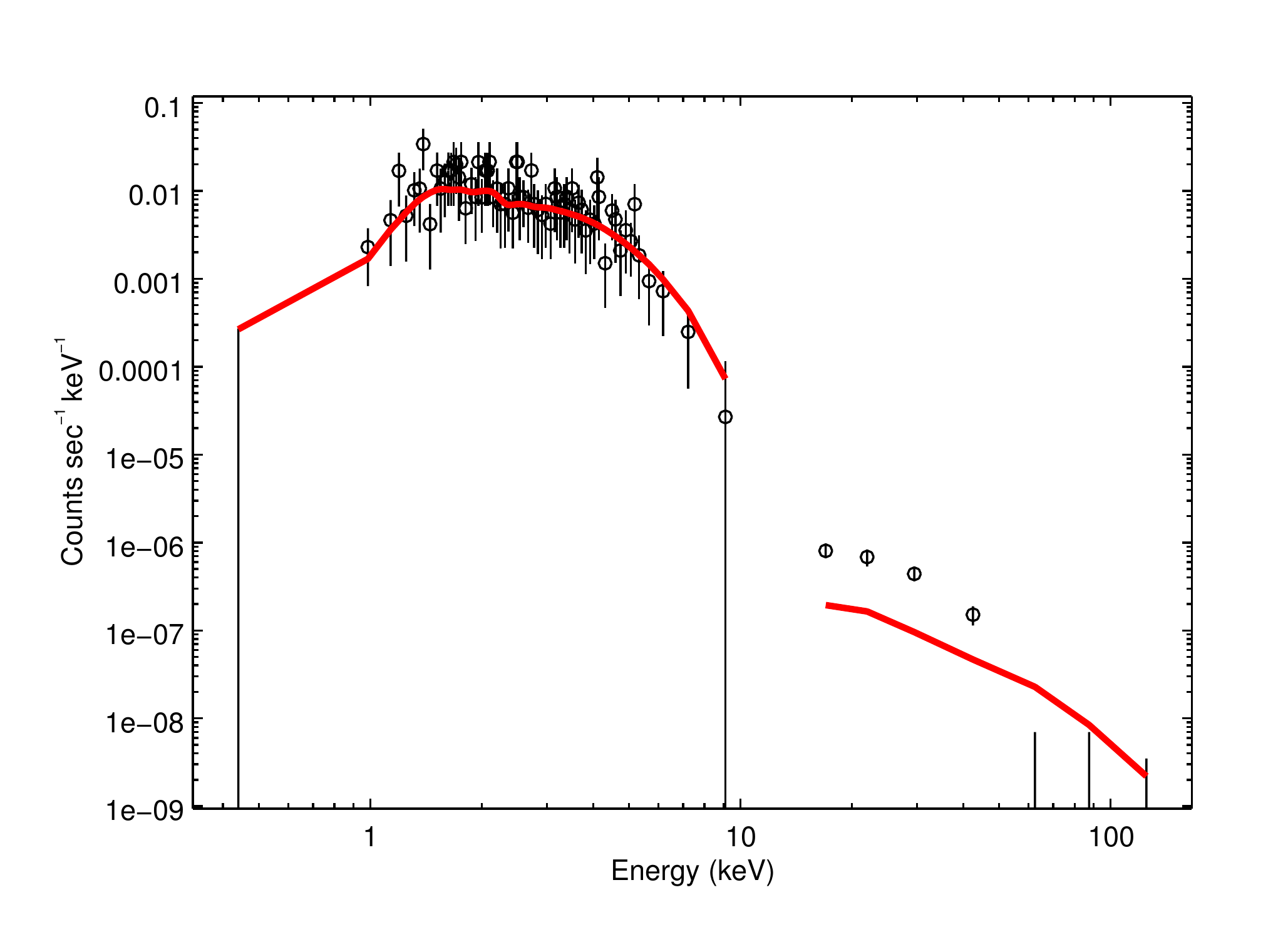} \label{fig-swift_spec}}
  \subfigure[\emph{Chandra} ACIS-S Spectra]{\includegraphics[width=0.5\textwidth]{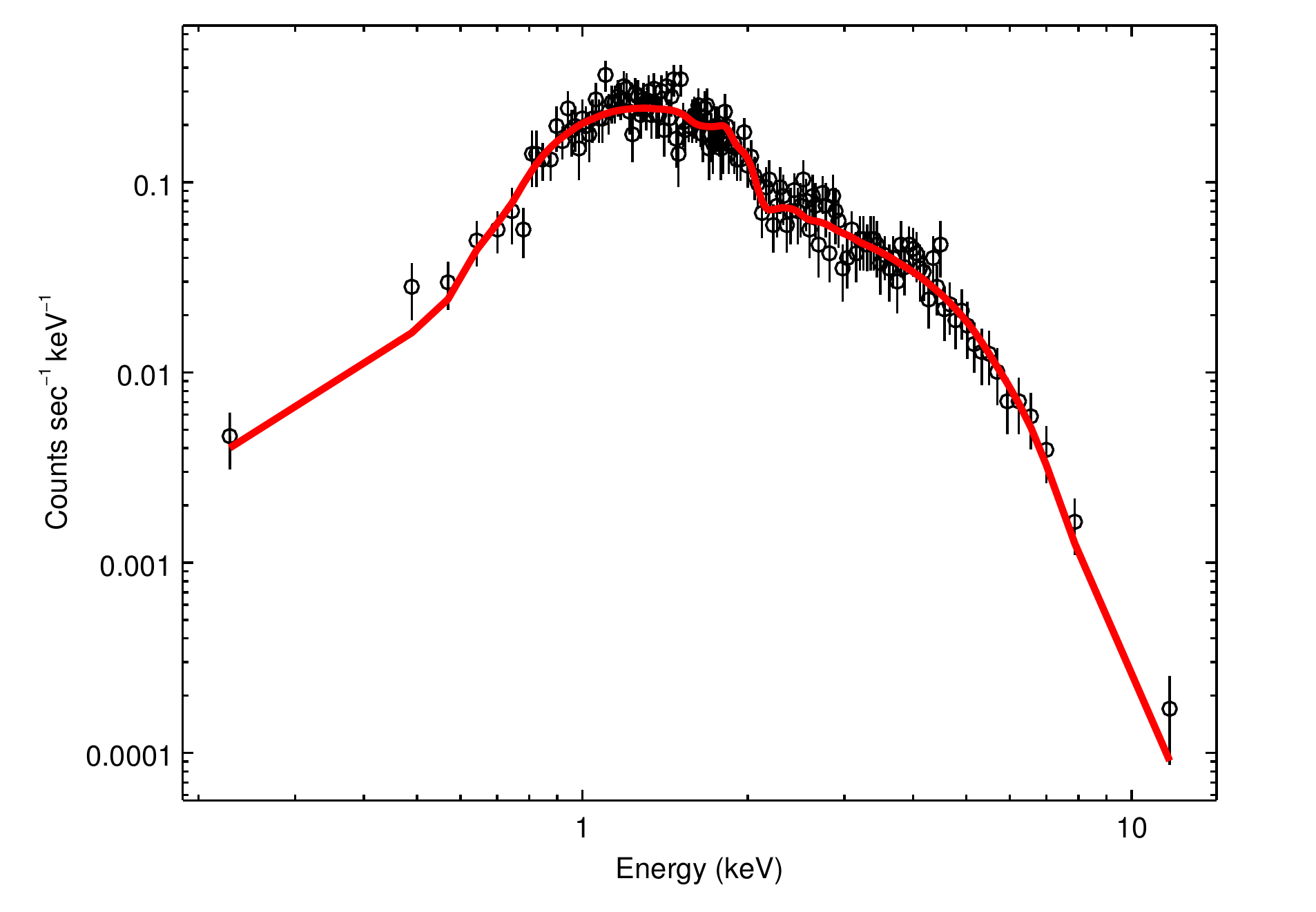} \label{fig-chandra_spec}}
\caption{The X-ray spectrum observed with the XRT and BAT on the \emph{Swift} observatory (a) and with \emph{Chandra} (b).  The data is shown as the black points with a five count binning.  The best spectral fit for each spectrum is shown as the red line.}
\label{fig-xray_spectra}
\end{figure}

With the X-ray detection of 2MASX J00423991+3017515 by the \emph{Swift} BAT and follow-up with \emph{Chandra}, we have two complementary sets of X-ray spectra.  The \emph{Swift} spectra consist of the lower energy $0.2-10$ keV XRT spectrum and the higher energy $14-145$ keV BAT spectrum.  While this covers a large energy range, it is poorer resolution.  The \emph{Chandra} spectrum taken with the ACIS-S detector covers a more limited energy range, only $0.2-10$ keV, but with higher resolution and better quality.  The photon counts for the XRT, BAT, and \emph{Chandra} spectra are $205$, $70$, and $2704$.  

The spectra from both telescopes were fit with Sherpa and preferred a simple absorbed powerlaw model,
\begin{equation}
M(E) = C\Big(E(1+z)\Big)^{-\Gamma}e^{-\eta_{H}\sigma(E(1+z))}
\end{equation} where the spectral model ($M$) depends on a constant value ($C$), the photon energy ($E$), the redshift ($z$) which is fixed at the redshift of the host galaxy, the column density of the intervening material ($\eta_H$), and the energy dependent photoelectric cross section ($\sigma$). The spectral data and fits are shown in Figure \ref{fig-xray_spectra}.  The best fitting parameters are given in Table \ref{table-xray_fits}.  The statistical fit of the \emph{Swift} spectrum was $\chi^{2}/DOF=98.3/45=2.1$ and for the \emph{Chandra} spectrum it is $\chi^{2}/DOF=82.7/146=0.57$.  Taking into account the intrinsic variability of accretion physics, both of the spectra obtained with the XRT and BAT on \emph{Swift}, and the ACIS-S instrument on \emph{Chandra} are similar.  

\citet{2017ApJS..233...17R} report a comprehensive analysis of the X-ray properties of the \emph{Swift} BAT 70-month AGN catalog that we can use for context in interpreting the properties of this AGN against its parent sample.  They find that the median column density is $\approx10^{22}$ N$_{H}$ cm$^{-2}$, and that the most obscured objects tend to be the least luminous.  Additionally, for nonblazar AGN, the typical power law index is $\Gamma=1.8$.  For the better constrained spectrum obtained with \emph{Chandra}, the spectral index is equivalent to the typical index of the BAT AGN, but the measured column density is on the lower end of the BAT AGN distribution.  Furthermore, the luminosity from 2-10 keV is $L_{2-10}=2.07\times10^{44}$ erg s$^{-1}$, high compared to the BAT AGN population.

\begin{table}
	\centering
	\caption{Best Fitting Parameters of X-ray Spectra}
	\label{table-xray_fits}
	\begin{tabular}{c|cc} 
		\hline
    Parameter & \emph{Swift} Spectrum  & \emph{Chandra} Spectrum \\
		\hline
    $C$ & $1.9\pm 0.6 \times 10^{-3}$ & $10.5\pm0.8\times10^{-4}$ \\
    $\Gamma$ & $2.22\pm0.09$ & $1.76\pm0.06$\\
    $\eta_{H}$ (atoms cm$^{-2}$) & $3.5\pm0.7\times10^{22}$ & $1.6\pm0.4\times10^{21}$\\
		\hline
	\end{tabular}
\end{table}

\begin{figure}
\centering
 \includegraphics[width=0.5\textwidth]{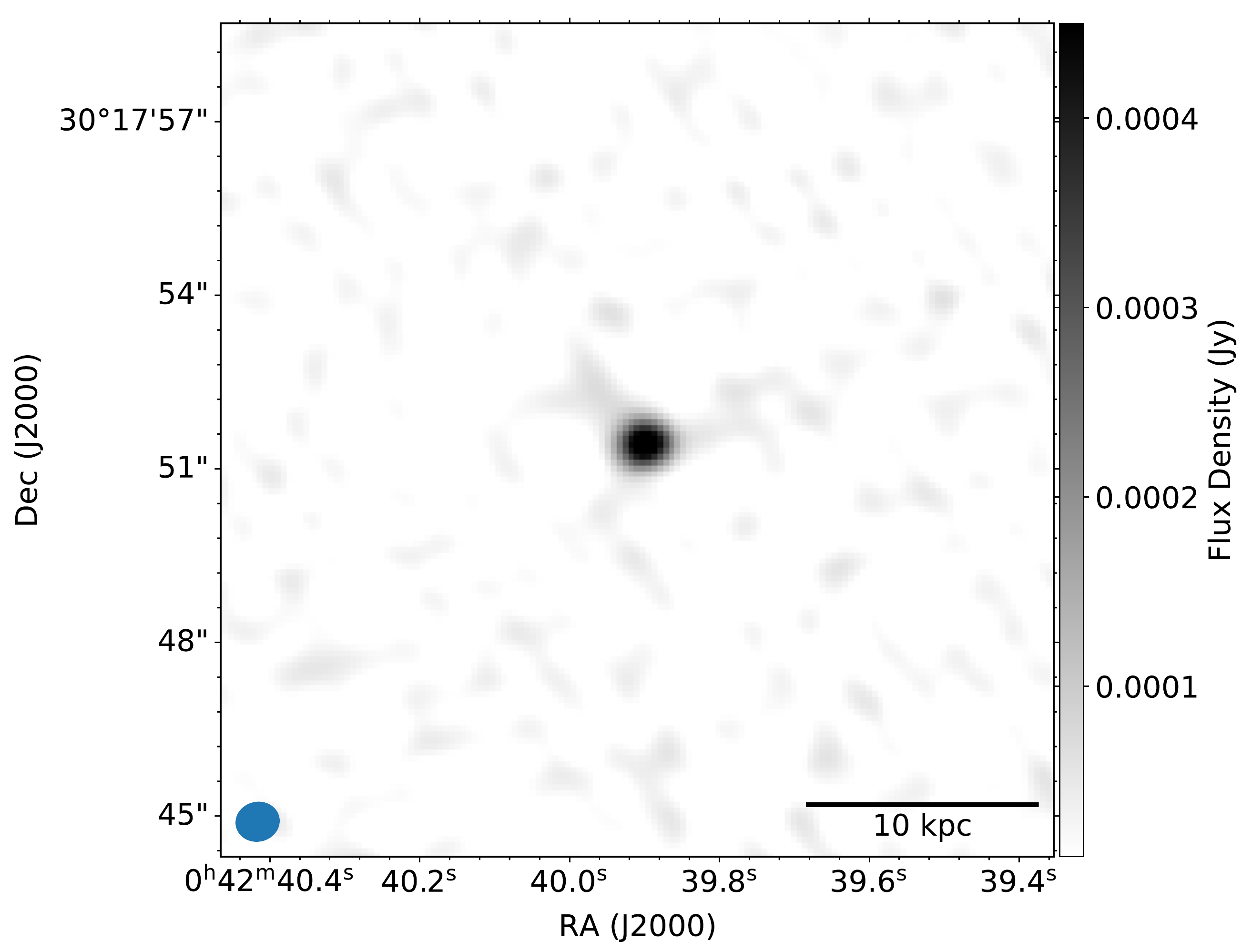}
\caption{JVLA 22~GHz radio image. A circle representative of the $\sim1$\arcsec\: beam is shown in blue.}
\label{fig-vla_image}
\end{figure}

In these spectra, any signature of a narrow Fe K$\alpha$ line at 6.4 keV is curiously absent from this data, even though it is a staple of AGN X-ray spectra \citep{1993ARA&A..31..717M, 1997MNRAS.286..513R}.  With a $90\%$ confidence, we can place an upper limit on the line strength of 111 eV, which is relatively weak.  Narrow Fe K$\alpha$ emission is created from the fluorescence of cold, optically thick iron atoms when the X-ray continuum of the AGN illuminates them.  This feature is often associated with the presence of the dusty torus invoked by the ``unified model" \citep{1993ARA&A..31..473A} because it has been determined to originate near the AGN but from an area of order $10\times$ further than the broad line emitting region \citep{2006MNRAS.368L..62N, 2011ApJ...738..147S}.

\begin{figure*}
\centering
 \includegraphics[width=\textwidth]{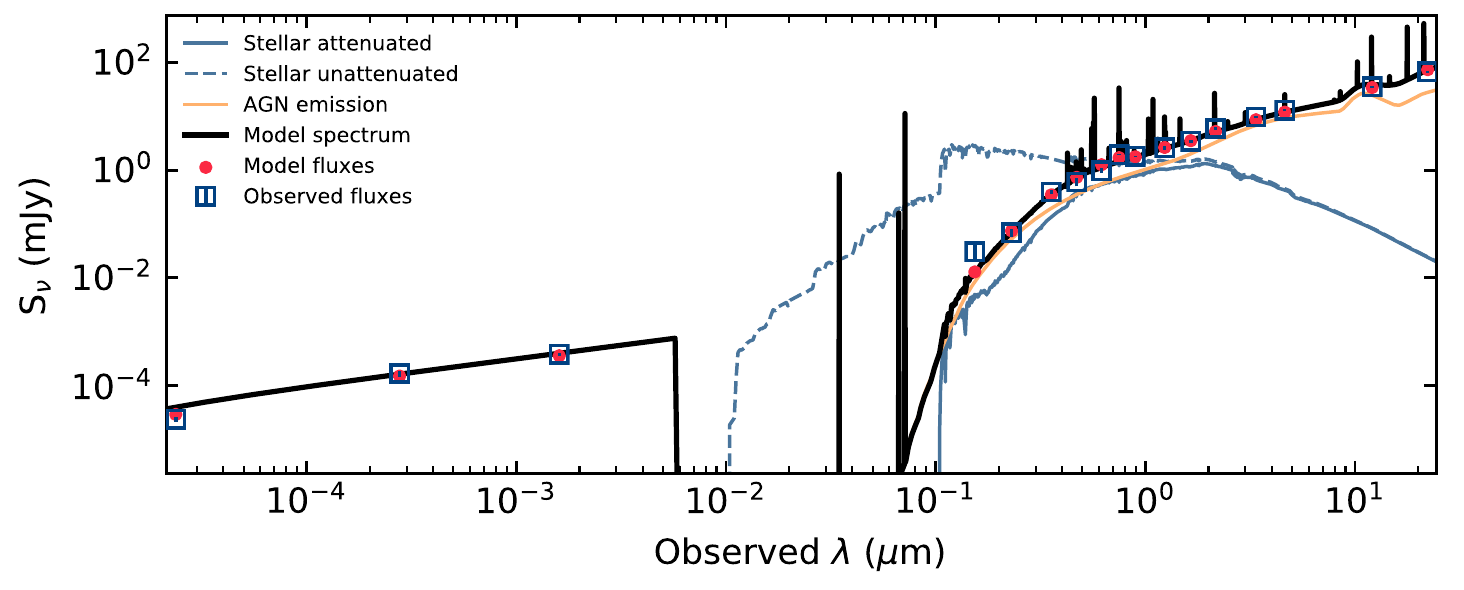}
\caption{The broadband SED and best fit model.  The data is shown as the blue boxes and the best fitting points are shown as red circles.  The individual model components shown are the unattenuated stellar component as the dashed blue line, the stellar component after attenuation from dust and gas as the solid blue line, the AGN component as the orange line, and the summed fitted model as the thicker black line.}
\label{fig-sed}
\end{figure*}

\subsection{JVLA Imaging}

The K-band JVLA image at $1\arcsec\:$ resolution is shown in Figure \ref{fig-vla_image}.  The image is consistent with an unresolved point source, with integrated flux density $S_\nu = 0.59$~mJy.   Contours of the image are overplotted on the HST image along with the \emph{Chandra} image contours in Figure \ref{fig-chandra_hubble_image}.  Like the \emph{Chandra} point source, the JVLA point source is centered on the AGN in the HST image.  The integrated flux density from the point source at 22 GHz yields a flux of $1.3\times10^{-16}$ erg s$^{-1}$ cm$^{-2}$ and a luminosity of $4.5\times10^{37}$ erg s$^{-1}$.

\subsection{Broadband Spectral Energy Distribution}

Closing our analysis, we present the broadband spectral energy distribution (SED) of 2MASX J00423991+3017515 from the far infrared to the hard X-ray, shown in Figure \ref{fig-sed}.  The data includes the J, H, and K bands of the Two Micron All Sky Survey \citep[2MASX,][]{2006AJ....131.1163S}; the $W1$, $W2$, $W3$, and $W4$ bands of the Wide-field and Infrared Survey Explorer \citep[WISE,][]{2010AJ....140.1868W} at 3.4, 4.6, 12, and 22 $\mu$m; the u, g, r, i, and z bands from the Sloan Digital Sky Survey \citep[SDSS,][]{2011ApJS..193...29A, 2015ApJS..219...12A}; the near-UV and far-UV from the Galaxy Evolution Explorer \citep[GALEX,][]{2005ApJ...619L...1M}; the 0.3-2 keV and 2-10 keV energy ranges from our \emph{Chandra} observations; and 14-195 keV energies from the \emph{Swift} BAT.  

The SED shows 2MASX J00423991+3017515 is luminous across all wavebands, but especially at the longer infrared wavelengths.  In fact, the IR luminosity across the $W1$, $W2$, $W3$, and $W4$ bands is $L_{W1} =4.6\times10^{45}$ ergs s$^{-1}$, $L_{W2} =4.6\times10^{45}$ ergs s$^{-1}$,  $L_{W3} =3.4\times10^{45}$ ergs s$^{-1}$, and $L_{W4} =4.9\times10^{45}$ ergs s$^{-1}$.  This places the galaxy in the regime of an ultra-luminous infrared galaxy (ULIRG), a unique class of galaxies that have high dust and gas fractions, as well as a copious star formation, AGN activity, or frequently a combination both.  

To constrain the emission mechanisms and glean insight into the galaxy properties, we utilized \texttt{X-CIGALE} \citep{2020MNRAS.491..740Y} which is an extension of the \texttt{CIGALE}\footnote{https://cigale.lam.fr} \citep{2005MNRAS.360.1413B, 2009A&A...507.1793N, 2019A&A...622A.103B} SED fitting package.  SED fitting can be difficult, especially for ULIRGs, because the complex interactions of dust heating from massive stars and AGNs, the subsequent absorption and emission of this energy, and the emission from the stellar population must all be handled properly.  \texttt{X-CIGALE} provides a self-consistent modeling framework by enforcing an energy balance between the different heating and cooling components, thus allowing for estimation of the star formation rate (SFR), stellar mass, and AGN contribution.  This is done by first building composite stellar populations from a simple stellar population (SSP) prescription based on a star formation history.  The nebular emission from gas ionized by massive stars along with emission from an AGN is calculated, and a specified law attenuates both the stellar population and this additional radiation.  The absorbed energy is then modeled as being re-emitted by dust in the mid-IR and far-IR.  Here, we used a delayed star formation history with the option for an exponential burst, the stellar initial mass function of \citet{2003ApJ...586L.133C}, the SSP model of \citet{2003MNRAS.344.1000B}, and the \citet{2012MNRAS.425.3094C} dust emission model.  We also used the \citet{2011MNRAS.415.2920I} model for the nebular emission and the \citet{2000ApJ...539..718C} law for reddening from dust extinction. The broadband AGN contribution is treated with the SKIRTOR model \citep{2012MNRAS.420.2756S, 2016MNRAS.458.2288S} which uses a two-phase, clumpy treatment of the torus.  The AGN is also allowed to be reddened by polar dust.

Given the lack of data at wavelengths longer than the WISE $W4$ filter at $22$ $\mu$m, we can't constrain the peak of the dust's black body emission, even though it is an important contributor to the overall model.  Therefore, we pin the value at a temperature of $T_{d}=40$ K, in line with values measured in samples of ULIRGs \citep{2010MNRAS.403..274C}.  The best fitting model to the SED has a $\chi^2/DOF=17.6/16=1.1$.  In the model, the main galaxy constituents, the AGN, dust, and stars, have luminosities of $L_{AGN}=(4.0\pm0.6)\times10^{11}\Lsun$, $L_{d}=(6.8\pm0.9)\times10^{11}\Lsun$, and $L_{s}=(7.7\pm1.1)\times10^{11}\Lsun$ with the stellar luminosity composed of $L_{s,o}=(1.6\pm0.9)\times10^{10}\Lsun$ in the old stellar population and $L_{s,y}=(7.5\pm1.0)\times10^{11}\Lsun$ in the young stellar population.  This translates into gas and stellar masses of $M_{g}=4\pm2\times10^{9}\Msun$ and $M_{s}=1.2\pm0.6\times10^{10}\Msun$.  The current measured SFR is elevated relative to the typical field galaxy at $SFR=150\pm50\:\Msun$ yr$^{-1}$, but inline with the ULIRG population.  The inferred star formation history suggests it recently underwent a burst of star formation and was higher in the recent past with an average SFR over 100 Myr of $SFR_{100}=22\pm7\:\Msun$ yr$^{-1}$, but a short term average over 10 Myr is significantly higher at $SFR_{10}=200\pm70\:\Msun$ yr$^{-1}$.

The AGN contribution to the SED originates from emission observed directly from the accretion disk and disk emission that has been intercepted by surrounding dust and then thermally reradiated.  The measured luminosities of these components in the SED fit are $L_{de}=(1.6\pm0.3)\times10^{11}\Lsun$ and $L_{therm}=(2.4\pm0.3)\times10^{11}\Lsun$, respectively.  The fit favors a face-on accretion disk (inclination angle $5\pm5$ degrees) embedded within a torus of opening angle $46\pm5$ degrees.

\section{Discussion}
\label{sec-discussion}

In this paper, we have presented a multi-wavelength spectroscopic and imaging study of an unusual AGN, 2MASX J00423991+3017515.  The source first stood out because of the striking $\sim$$1500$ km s$^{-1}$ kinematic offset in the peaks of every observable broad Balmer line from the corresponding narrow line emission, and additional spectroscopic monitoring confirmed that it was persistent.  This motivated high-resolution, multi-band imaging with HST, \emph{Chandra} and the JVLA.  The radio, NIR, optical, UV, and X-ray images show a single AGN with an apparent spatial displacement of $\sim4$ kpc from the center of the galaxy's disk, as identified from the GALFIT model of the HST image.  The X-ray spectra from \emph{Swift} and \emph{Chandra} revealed a mundane X-ray spectrum, with the caveat that the Fe K$\alpha$ line is weak, or possibly missing, and the column density to the X-ray emitter is low.  SED fitting with archival survey data found that the AGN hosting galaxy is a ULIRG that is actively forming stars.  Based on our analysis of these initial data, it is clear that this galaxy has an active and ongoing merger history and represents a rare phase of galaxy and SMBH evolution. Now, we will examine several possible scenarios that might explain the unusual features of this object.  As a starting point, we focus on the apparent spatial displacement of the AGN and assume it could have arisen from either the interaction with the companion, the presence of multiple SMBHs in the galaxy, or from the actual movement of the AGN itself.

Given how close the faint companion is to the AGN hosting galaxy, it is reasonable to expect that a strong interaction between the two galaxies disturbed their morphologies.  However, without knowing the dynamics, we can only speculate as to how the interaction has developed. For example, we could imagine the companion hit the primary with a strong glancing blow that partially disrupted its disk, or that the interaction elongated the primary in such a way that from our vantage point it looks as if the nucleus is displaced.  If a violent collision were capable of stripping away a significant portion of the stellar disk--the feasibility of which is unclear--then we would need to assume that the resulting debris field is dispersed to the point that it is unobservable in our HST data. Alternatively, if the galaxy interaction may have created a tidal tail or stretching of the primary galaxy along our line of sight, the apparent displacement may simply be from an unusual chance projection. 

Physically separated and displaced AGNs are also observed in kpc-scale dual and triple AGN systems following a major merger before the SMBHs have settled to the bottom of the gravitational potential well and formed a close binary.  Dozens of them have been found individually \citep[e.g.,][]{2003ApJ...582L..15K, 2018ApJ...862...29L, 2018MNRAS.478.1326L, 2019ApJ...883..167P} and through dedicated searches \citep[e.g.,][]{2011ApJ...737..101L, 2011MNRAS.418.2043E, 2012ApJ...746L..22K, 2013ApJ...777...64C, 2017ApJ...848..126S}, across the electromagnetic spectrum.  If in these types of systems the additional AGNs were either quiescent or accreting but heavily obscured by dust and gas and therefore not visible, they could appear as a system with a lone AGN that is spatially displaced by several kpc, like the system studied here.  However, we find no indication of additional SMBHs in any of our HST, \emph{Chandra}, or JVLA imaging.  The lack of additional AGNs in the JVLA imaging is particularly telling since \citet{2020MNRAS.492.4216S} found that similarly configured 22 GHz observations of 100 AGNs in the BAT AGN Spectroscopic Survey \citep{2017ApJ...850...74K} produced a $96\%$ radio core detection rate, and even very obscured AGN at low masses ($10^6$) have been seen in radio continuum searches.

Both of these scenarios can explain the offset nucleus, but would require an independent explanation for the large offsets observed in the broad Balmer lines. \citet{2010ApJS..187..416L} showed that if the broad line region is large and there are features like spiral arms, offset peaks can be observed in broad line profiles.  Further, if the BLR gas is distributed asymmetrically in clumpy overdensities, this can translate into an asymmetric line profile.  Additionally, outflowing winds have also been identified as sources of offset broad line peaks, depending on the viewing angle and other properties \citep[e.g.,][]{2010MNRAS.409.1033M, 2017PASA...34...42Y}.

A surprising cause of the BLR peak offset could be a binary SMBH system embedded within the larger merging system.  Binary AGNs occur at smaller separations when two SMBHs are gravitationally bound, a scale that would be unresolved in our HST imaging.  If only one SMBH is actively accreting, the binary system could appear as an AGN with a large velocity offset in its broad line emission, owing to its bulk orbital motion.  Indeed, several AGN have been discovered with very large offsets in their broad line emission that are suggestive of a sub-pc binary, but the velocities measured from the broad Balmer line offsets are higher than we find \citep[$3000$ km s$^{-1}$;][]{2009ApJ...707..936S,  2009Natur.458...53B, 2009MNRAS.398L..73D, 2009ApJ...697..288B, 2012ApJS..201...23E}.  Even so, \citet{2014ApJ...789..140L} used changes in the peaks of velocity shifted broad line peaks to search these these types of systems where only one SMBH in the sub-pc binary is accreting.

If we consider that the AGN itself could be in motion, a slingshot recoil or a GW recoil could potentially offer a single explanation for both the spatial displacement and velocity offset together, but these are two exotic scenarios.  A slingshot recoil would also imply a second, unseen SMBH is in the galaxy as in the dual AGN case, but at much smaller, $\sim$mpc separations. It is also possible that a three-body interaction leading to a slingshot recoil of one SMBH would shrink the orbit of the remaining binary SMBH enough to cause a rapid merger \citep[e.g.,][]{2016MNRAS.461.4419B}.  Nevertheless, objects like CID-42 \citep{2010ApJ...717..209C, 10.1093/mnras/sts114, 2012ApJ...752...49C}, SDSSJ092712.65+294344.0 \citep{2008ApJ...678L..81K}, CXO J101527.2+625911 \citep{2017ApJ...840...71K}, and 3C 186 \citep{2017A&A...600A..57C, 2018ApJ...861...56C}, among others, whose features largely resemble those of 2MASX J00423991+3017515 have been found and could be examples of this same phenomenon.

With our current data, no firm conclusions can be drawn about the origin of 2MASX J00423991+3017515's odd character, but we can broadly assess the plausibility of the different cases.  First, we can be fairly confident the velocity shift in the broad Balmer lines isn't caused by a close SMBH binary with only one actively accreting SMBH because the broad line offset didn't show any statistically significant shift and the measured velocity shift is lower than that expected from such a binary.  Plus, such  systems are scarce, so finding one within an unusual merger or a dual AGN system would be unlikely.  Second, there is no evidence for additional SMBHs with any resolvable separation, so attributing the spatial displacement to a multi-SMBH system where only one is seen as an AGN is also disfavored. It is likely that if additional SMBHs are present in the system they would be accreting because of the plentiful gas supply in the host galaxy of 2MASX J00423991+3017515 and its interaction with the companion. The \emph{Chandra} imaging probed to a depth of $4\times10^{41}$ erg s$^{-1}$ at $2-10$ keV which would have captured any other typical Seyfert-like AGNs yet none were found.  Perhaps there may be quiescent SMBHs, low-luminosity AGNs (LLAGNs), or heavily obscured AGNs, but currently there are no indications they exist.

Therefore, the two remaining explanations for this object's peculiar properties are that it is an AGN with an unusually strong outflowing wind residing in a merger that is uniquely oriented to give the appearance of a spatial offset, or it is a recoiling AGN.  The former case is appealing because there is clearly an interaction with its companion and the AGN is located in a bulge-like region of the galaxy.  The outflowing wind needed to cause the offset broad lines would lie on the extreme upper end of observed velocities and be relatively rare, though not unprecedented.  \citet{2012ApJS..201...23E} conducted a systematic search for offset broad H$\beta$ lines with a sample of 15,900 QSOs from the Sloan Digital Sky Survey and found 88 objects that had offsets larger than 1000 km s$^{-1}$ with only 31 having offsets larger than 1500 km s$^{-1}$.  It is hard to quantify the probability of such a wind occurring simultaneously yet independently in a unique merger, but it should be low.  One benefit of the recoil scenario is that it could explain the spatial displacement of the AGN from the apparent center of the host galaxy and the kinematic offset in the broad Balmer line emission without having to appeal to undetected SMBHs or winds. In both a slingshot and GW recoil, however, a prior merger or mergers must be invoked to create a close binary or triple SMBH system. A more detailed understanding of the galaxy's dynamics is needed to understand why there would be a large and lopsided enhancement of stars and star formation.  In this case, the current ongoing merger would be unrelated to the SMBH kick, implying a very active merger history for this galaxy.

2MASX J00423991+3017515 is indeed puzzling, and additional follow-ups are needed to distinguish between the possible scenarios.  Thankfully, its proximity will enable high-quality targeted observations, which we are pursuing to determine whether it is undergoing an unusual merger or if it might be a \emph{bona fide} recoiling AGN.  Using ALMA to leverage the abundance of settled dust and gas in the primary galaxy to map the rotation curve of galaxy and determine the dynamical center of the galaxy is one of the highest priority observations.  Knowing the AGN's position relative to the dynamical center will elucidate if it is actually displaced from the bottom of the gravitational potential. UV spectra could clarify the source of the broad line shift by exposing features like blue-shifted absorption lines that would support the wind explanation or offset broad lines in other atomic species which would support the recoil scenario.  Deep high-resolution X-ray and radio imaging studies are also needed to place tighter limits on the presence additional AGNs in the system to ensure there are no low-luminosity LLAGNs or obscured AGNs.  Deeper X-ray spectral observations will aid in determining the strength of the Fe K$\alpha$ line to understanding if a dusty torus surrounds this AGN. A recoiling SMBH shouldn't retain a torus of molecular gas after being kicked, so the absence or weakness of the Fe K$\alpha$ line in the X-ray spectrum could be an indication the it is missing or at least weakened.  However, this must be reconciled with the SED which was best fit by face-on accretion disk surrounded by hot dust, presumably heated by the AGN.  This should be an ideal configuration to observe the Fe K$\alpha$ line.  IR studies of the hot dust could relieve this possible conflict by mapping the distribution of the hot dust.  For instance, is it localized near the AGN or spread throughout the galaxy in some other manner?

The recoiling AGN hypothesis is particular interesting, and if does withstand further vetting, this discovery would have several important astrophysical implications for the nature of galaxy and SMBH coevolution.  First, and most obviously, it would demonstrate that SMBHs are coalescing, and it would deliver a key link between theory and observation.  The confirmed detection of a GW recoiling AGN would also demonstrate that SMBH spin alignment does not always operate efficiently even in a gas rich galaxy, at odds with the expectation that spin alignment may occur rapidly if the SMBHs are embedded in a circumbinary gas disk \citep[e.g.,][]{2007ApJ...661L.147B, 2009MNRAS.396.1640D, Miller_2013}.  Theoretical distributions of kick velocities have been predicted based on fits to numerical relativity simulations \cite[e.g.,][]{2007ApJ...668.1140B, 2012PhRvD..85h4015L}.  Lower velocity kicks are produced from coalesce events of SMBHs with perfectly aligned spins ($< 200$ km s$^{-1}$) and spins aligned within a few degrees ($\le 600$ km s$^{-1}$).  At the extreme upper end, kicks could reach velocities as high as $5000$ km s$^{-1}$ \citep{2007PhRvL..98w1102C} and potentially eject the newly merged SMBH from the galaxy entirely.  \citet{2010PhRvD..81h4023L} found that large kicks should be rare, but $~25\%$ should be higher than $1000$ km s$^{-1}$ if spins are randomly oriented.  When observational effects and spatial and velocity resolution constraints are also considered, they predict roughly $2\%$ of recoiling AGNs are expected to have measurable velocities along our line-of-sight of greater than $1000$ km s$^{-1}$.  The kick velocity of this AGN as inferred from the broad line offsets in the GW recoil scenario would require misaligned spins at coalescence.

If, on the other hand, it is a slingshot recoil, it would suggest that stalled binaries are common enough for triple SMBH systems to be dynamically important.  The prevalence of slingshot recoil kicks is not known, but recent theoretical work indicates that they may be relatively common \citep{2017MNRAS.464.3131K, 2018MNRAS.477.3910B}.

\section{Conclusions}
\label{sec-conc}

In this paper, we have presented a multiwavelength imaging and spectroscopic study of 2MASX J00423991+3017515.  The system is composed of an AGN in an edge-on disk galaxy that is interacting with a nearby companion.  2MASX J00423991+3017515 has the following unique properties:\begin{itemize}
\item{The AGN, observed in the radio, IR, optical, UV, and X-ray bands, is found to be spatially displaced from the center of the host galaxy's disk component by 3.8 kpc in a GALFIT model of the system.}
\item{In optical spectra from two different epochs, all of the Doppler broadened Balmer emission (H$\alpha$, H$\beta$, \& H$\gamma$) is consistently blue-shifted from the narrow line emission by $1540$ km s$^{-1}$.  Furthermore, we localized the source of the velocity offset broad H$\alpha$ line to be the spatially displaced AGN.}
\item{The X-ray spectrum lacks any sign of Fe K$\alpha$ emission with an upper limit on the equivalent width of the line of 111 eV.}
\end{itemize}
We explore several possibilities to explain the peculiar features of this system and propose that they could be because the AGN in 2MASX J00423991+3017515 may be in the midst of a merger while simultaneously showing a strong outflowing wind, or it may be recoiling from either a gravitational ``slingshot" from a three-body SMBH interaction or from the asymmetric emission of GWs following the coalescence of two progenitor SMBHs.  Rigorous follow-up studies will be the only way to uncover the true nature of 2MASX J00423991+3017515.  The most pressing observations to acquire are a high-fidelity kinematic mapping of the dust and gas in the system to constrain the dynamics, UV spectra to search for wind signatures, and high-quality deep radio and X-ray imaging to search for additional hidden AGNs.

\section*{Acknowledgements}

We thank the anonymous referees for thorough and helpful reviews of this manuscript.  The authors thank Chien Y. Peng for his immense help with the GALFIT modeling and Richard Cosentino for feedback on early versions of the paper.  This work was supported in part by NASA through two grants: HST-GO-14732 from the Space Telescope Science Institute and Chandra Award Number G07-18115X issued by the \emph{Chandra} X-ray Observatory Center.  LB acknowledges support from NSF grant AST-1909933.  The optical spectra were obtained using University of Colorado time at the Apache Point Observatory which is operated by the Astrophysics Research Consortium (ARC) and University of Maryland time on the Lowell Discovery Telescope (LDT) at Lowell Observatory. Lowell is a private, non-profit institution dedicated to astrophysical research and public appreciation of astronomy and operates the LDT in partnership with Boston University, the University of Maryland, the University of Toledo, Northern Arizona University and Yale University.

\section*{Data Availability}
The data underlying this article will be shared on reasonable request to the corresponding author.

\bibliography{manuscript}

\newpage

\appendix
\section{UT 2016-12-05 Systematic Uncertainties}
\label{sec-ldt_systematics}

Variability is an inherent property of the accretion processes that power AGNs.  This could naturally explain the changes we observed in the optical spectrum of 2MASX J00423991+3017515 from UT 2011-08-07 to UT 2016-12-05.  However, there are indications that data acquisition and/or reduction issues could be the root cause of the observed change in the UT 2016-12-05 spectrum.  In particular, the blue spectrum taken on UT 2011-08-07 has a continuum that strongly increases at shorter wavelengths, expected from a Type 1 AGN, while the UT 2016-12-05 data shows a suppressed continuum.  Caution is needed in interpreting this and the change in the blueward wing of the H$\beta$ broad line as true variability, though, because additional systematic uncertainties could be caused by the poor observing conditions during the data collection.  

The data reduction was run several times to ensure it was executed properly.  Particular attention was given to the flat fielding and detector sensitivity calibration steps since they could artificially flatten the spectrum if done improperly.  Neither the response curve used for flat fielding nor the sensitivity curve used for flux calibration showed anything unusual.  Visually, the points used for fitting the response curve showed a smooth curve with no anomalous features (like peaks, sharp deviations, undulations, etc.) that might indicate anything unusual.  No large structure was present in the residuals and their RMS was $0.04$ with the maximum deviation being $0.15$ for a single spurious point.  Given the baseline value of $\sim37.5$, this represents a typical residual of $\sim0.1\%$ and a max of $\sim0.5\%$ for the one point.

On both UT 2011-08-07 and UT 2016-12-05 the data was acquired with the slit aligned along the galaxy's disk in an attempt to capture the maximum amount of stellar light, rather than at the parallactic angle.  This strategy can lead to lost flux from differential atmospheric refraction; effects that are more pronounced at bluer wavelengths and when observations are taken at higher airmasses.  To mitigate this, the three exposures in the blue set of observations taken on UT 2016-12-05 were made at airmasses of 1.0, 1.01, and 1.04.  \citet{1982PASP...94..715F} presents guidance on the expected impact from differential atmospheric refraction and given these low airmasses and our $1.5\arcsec$ slit, the effect should be small.  Still, even though the standard stars were observed with the slit at the same position, changes in the atmospheric turbulence that reduced the seeing to $1.5\arcsec$ could have exacerbated the differential refraction and caused more light to fall outside the slit, thereby diminishing the captured flux.

\bsp	

\label{lastpage}
\end{document}